\documentclass[10pt]{article}
\newtheorem{theorem}{Theorem} 
\newtheorem{lemma}[theorem]{Lemma}

\newtheorem{definition}[theorem]{Definition}
\usepackage{cite}
\usepackage{amsmath}
\usepackage{amssymb}
\usepackage{psfig}
\usepackage{epsfig}
\usepackage{color}

\usepackage{setspace}

\newcommand{\tctla}{\mbox{TCTL$^\forall$}}
\newcommand{\tctle}{\mbox{TCTL$^\exists$}}

\newcommand{\true}{\mbox{\em true}}
\newcommand{\false}{\mbox{\em false}}

\newcommand{\breach}{\mbox{\tt reachable-bck}}

\newcommand{\evalg}{\mbox{\tt Eval-EDGF}}
\newcommand{\eval}{\mbox{\tt Eval}}
\newcommand{\hst}{\\\hspace*{4mm}}
\newcommand{\hstt}{\\\hspace*{8mm}}
\newcommand{\hsttt}{\\\hspace*{12mm}}

\newcommand{\pf}{\noindent\mbox{\bf Proof : }}

\newcommand{\pfrr}{\Box}
\newcommand{\until}{\mbox{$\cal U$}}
\newcommand{\pevt}{\Diamond}

\def\qed{\ifmmode\|\else{\unskip\nobreak\hfil
\penalty50\hskip1em\null\nobreak\hfil$\blacksquare$
\parfillskip=0pt\finalhyphendemerits=0\endgraf}\fi}

\newenvironment{list1}{\begin{list}{$\bullet$}
{\topsep 0 pt \parsep 0 pt \partopsep 0 pt \itemsep 0 pt}}{\end{list}}
\newenvironment{list2}{\begin{list}{$-$}
{\topsep 0 pt \parsep 0 pt \partopsep 0 pt \itemsep 0 pt}}{\end{list}}

\newcounter{cabbage1}

\newcounter{cabbage2}

\newcounter{cabbage3}

\newcounter{bean1}

\newcounter{bean2}

\newcounter{bean3}

\newcounter{bean4}

\newcounter{bean5}

\newcounter{bean6}

\begin{document}

\title{TCTL Inevitability Analysis of Dense-time Systems\thanks{The
work is partially supported by NSC, Taiwan, ROC under
grants NSC 90-2213-E-002-131, NSC 90-2213-E-002-132, and by the Internet protocol verification
project of Institute of Applied Science \& Engineering Research, Academia Sinica, 2001.
}}
\author{Farn Wang \\
Dept. of Electrical Engineering, National Taiwan University\\
Taipei, Taiwan 106, Republic of China \\
+886-2-23635251 ext. 435; FAX +886-2-23691707; farn@cc.ee.ntu.edu.tw\\[3mm]
Geng-Dian Hwang, Fang Yu\\
Institute of Information Science, Academia Sinica, Taipei, Taiwan 115, ROC\\[3mm]
Tool available at: http://cc.ee.ntu.edu.tw/\~{ }val/red
        }

\date{}
\baselineskip 10pt
\thispagestyle{empty}
\pagestyle{plain}
\maketitle

\begin{abstract}
Inevitability properties in branching temporal logics are
of the syntax $\forall\pevt \phi$, 
where $\phi$ is an arbitrary (timed) CTL formula.
In the sense that "good things will happen", they are parallel to the "liveness"
properties in linear temporal logics.
Such inevitability properties in dense-time logics can be analyzed with greatest fixpoint
calculation.
We present algorithms to model-check inevitability properties 
both with and without requirement of non-Zeno computations.
We discuss a technique for 
early decision on greatest fixpoints in the temporal logics.  
Our algorithms come with a $d$-parameter for the measurement of time-progress.  
We have experimented with various issues, which may affect the 
performance of TCTL inevitability analysis.  
Specifically, we report the performance of our implementation 
w.r.t. various $d$-parameter values and  
with or without the non-Zeno computation requirement in the evaluation 
of greatest fixpoints.
We have also experimented with safe abstration techniques for model-checking 
TCTL inevitability properties.
Analysis on the experiment data 
helps clarify 
how various techniques can be used to improve verification of inevitability properties.  
\end{abstract}



\section{Introduction}

In the research of verification, very often two types 
of specification properties attract most interest from academia and industry. 
The first type specifies that {\em "bad things will never happen"}   
while the second type specifies that {\em "good things will happen"}\cite{AS85}.    
In linear temporal logics, the former is captured by modal operator $\pfrr$
while the latter by $\pevt$\cite{Pnueli77}.  
Tremendous research effort has been devoted to the efficient analysis of 
these two types of properties in the framework of linear temporal logics\cite{MP95}.  
In the branching temporal logics of (timed) {\em CTL}\cite{ACD90,CE81,CES86}, 
these two types can be mapped to 
modal operators $\forall\pfrr$ and $\forall\pevt$ respectively.  
$\forall\pfrr$ properties are called {\em safety} properties
while $\forall\pevt$'s are usually called {\em inevitability} 
properties\cite{Emerson87,MOP89}.  
In the domain of dense-time system verification, 
people have focused on the efficient analysis of 
safety properties\cite{DOTY96,HW99,MLAH99a,MLAH99b,PL00,Wang01a,Wang01b,Wang03,WH98,WH02,Yovine97}.  
Inevitability properties in {\em Timed CTL (TCTL)}\cite{ACD90,HNSY92} are comparatively more complex to analyze due to 
the following reason.  
In the framework of model-checking, to analyze an inevitability 
property, say $\forall\pevt \phi$, we actually compute the
set of states that satisfy the negation of inevitability, 
in symbols $[[\exists\pfrr\neg\phi]]$, 
and then use the intersection emptiness
between $[[\exists\pfrr\neg\phi]]$ and the initial states for the answer to the 
inevitability anlaysis.  
However, property $\exists\pfrr\neg \phi$ in TCTL semantics  
is only satisfied with non-Zeno computations\cite{ACD90}.   
(Zeno computations are those counter-intuitive infinite computations
whose execution times converge to a finite value\cite{HNSY92}.)  
For example, a specification like
\begin{center}
"along all computations, eventually a bus collision will happen in
three time units"
\end{center}
can be violated by a Zeno computation whose execution time converges to 
a finite timepoint, e.g. 2.9.
Such requirement on non-Zeno computations may add complexity to the 
evaluation of inevitability properties. 
In this work, we present our symbolic TCTL model-checking algorithm 
which can handle the non-Zeno requirement in the
evaluation of {\em greatest fixpoints}. 
The evaluation of inevitability properties in TCTL 
involves nested reachability analysis 
and demands much higher complexity than simple safety analysis.

To contain the complexity of TCTL inevitability, it is important 
to integrate new and old techniques for a performance solution.  
In this paper, we investigate three approaches.  
In the first approach, we investigate how to adjust a parameter value 
in our greatest fixpoint evaluation algorithms for better performance.  
We have carried out experiments to get insight on this issue.  

In the second approach, we present a technique called
{\em Early Decision on the Greatest Fixpoint (EDGF)}.
The idea is that, in the evaluation of the greatest fixpoints,
we start with a state-space and iteratively pare states from it until we reach
a fixpoint.
Throughout iterations of the greatest fixpoint evaluations, the state-space is 
non-increasing.
Thus, if in a middle greatest fixpoint evaluation iteration,
we find that target states have already been pared from the greatest fixpoint,
we can conclude that it is not possible
to include these target states in the fixpoint.
Through this technique, we can reduce time-complexity irrelevant to the
answer of the model-checking.
As reported in section~\ref{sec.experiments}, 
significant performance improvement has been observed in several benchmarks.  

Our third approach is to use abstraction techniques\cite{CC77,CC92a,WongToi95}.  
We shall focus on a special subclass, \tctla, 
of TCTL in which every formula
can be analyzed with safe abstraction if over-approximation is used in the 
evaluation of its negation.  
For example, we may write the following formula in the subclass.
\begin{center}
$\forall\pfrr \left(\mbox{\tt request} 
	\rightarrow \forall \pfrr\left(\mbox{\tt service}
		\rightarrow \forall \pevt {\tt request}
	\right)
\right)
$
\end{center} 
This formula says that if a request is responded by a service, 
then a request will follow the service.  
This subclass allows for nested modal formulas and 
we feel that it captures many TCTL inevitability properties.

One challenge in designing safe abstraction techniques in model-checking 
is making them accurate enough to discern many true properties, 
while still allowing us to enhance verification performance.  
In previous research, people have designed many abstraction techniques 
for reachability analysis\cite{Balarin96,DOTY96,PL00,WongToi95,Yovine97}, as have we\cite{WHY03}.  
However, for model-checking formulas in \tctla, abstraction accuracy can be a 
bigger issue since the inaccuracy in abstraction can be potentially magnified when we 
use inaccurate evaluation results of nested modal subformulas to evaluate 
nesting modal subformulas with abstraction techniques.  
Thus it is important to discern accuracy of previous abstraction techniques 
in discerning true \tctla formulas.  

In this paper, we also discuss another possibility for abstract evaluation 
of greatest fixpoints, which is to omit the requirement for non-Zeno computations 
in TCTL semantics. 
As reported in section~\ref{sec.experiments}, many benchmarks are true 
even without exclusion of Zeno computations.  

Finally, we have implemented these ideas in our model-checker/simulator 
{\tt red} 4.1\cite{Wang03}.
We report here our experiments to observe the effects of parameter values, EDGF, 
various abstraction techniques, and 
non-Zeno requirements on our inevitability analysis.  
We also compare our analysis with Kronos 5.1\cite{DOTY96,Yovine97}, 
which is another model-checker for full TCTL.  

Our presentation is ordered as follows.  
Section~\ref{sec.relwork} discusses several related works.  
Sections~\ref{sec.system} and \ref{sec.tctl} give 
brief presentations of our model, {\em timed automata (TA)}, and TCTL.  
Section~\ref{sec.algorithm} presents our TCTL model-checking algorithm 
with requirements for non-Zeno computations.  
Section~\ref{sec.edgf} improves our model-checking algorithm using an 
EDGF technique.  
Section~\ref{sec.Zeno} gives another version of a greatest fixpoint 
evaluation algorithm by omitting the requirement of non-Zeno computations.  
Section~\ref{sec.tctle} identifies the subclass \tctla of TCTL 
which supersedes many inevitability properties, 
while allowing for safe abstract model-checking by using over-approximation techniques.  
Section~\ref{sec.experiments} illustrates our experiment results and helps clarify 
how various techniques can be used to improve analysis of inevitability properties.  
Section~\ref{sec.conc} is the conclusion.

\section{Related work\label{sec.relwork}} 

The TA model with dense-time clocks was first presented in \cite{AD89}.  
Notably, the data-structure of DBM is proposed in \cite{Dill89} for the 
representation of convex state-spaces of TA. 
The theory and algorithm of TCTL model-checking were first given in \cite{ACD90}.  
The algorithm is based on region graphs and helps manifest the 
PSPACE-complexity of the TCTL model-checking problem.  

In \cite{HNSY92}, Henzinger et al 
proposed an efficient symbolic model-checking algorithm for TCTL.  
However, the algorithm does not distinguish between Zeno and non-Zeno computations.  
Instead, the authors proposed to modify TAs with Zeno computations to 
ones without.  
In comparison, our greatest fixpoint evaluation algorithm is innately able 
to quantify over non-Zeno computations.  

Several verification tools for TA 
have been devised and implemented so far\cite{DOTY96,HW99,MLAH99a,MLAH99b,PL00,Wang01a,Wang01b,Wang03,WH98,WH02,Yovine97}.
UPPAAL\cite{BLLPW96,PL00} is one of the popular tool with DBM technology. 
It supports safety (reachability) analysis in forward reasoning techniques.  
Various state-space abstraction techniques and compact representation 
techniques have been developed\cite{BLPWW99,LLPY98}.   
Recently, Moller has used UPPAAL with abstraction techniques
to analyze restricted inevitability properties with no 
modal-formula nesting\cite{Moller02}.  
The idea is to make model augmentations to speed up the verification performance. 
Moller also shows how to extend the idea to analyze TCTL with only universal 
quantifications.  
However, no experiment has been reported on the verification of nested modal-formulas.  

Kronos\cite{DOTY96,Yovine97} is a full TCTL model-checker with DBM technology and 
both forward and backward reasoning capability.  
Experiments to demonstrate how to use Kronos to verify 
several TCTL {\em bounded inevitability} properties is demonstrated in \cite{Yovine97}. 
({\em Bouonded inevitabilities} are those inevitabilities specified with a deadline.)  
But no report has been made on how to enhance the performance of general inevitability 
analysis.  
In comparison, we have discussed techniques like EDGF and abstractions which
handle both bounded and unbounded inevitabilities.  

DDD is a reachability analyzer based on BDD-like data-structures  
for TA\cite{MLAH99a,MLAH99b}.  

SGM\cite{WH02} is a compositional safety (reachability) analyzer for TA, 
also based on DBM technology. 
A newer version also supports partial TCTL model-checking.  

CMC is another compositional model-checker\cite{LL98}. 
Its specification language is a restricted subclass of $T_\mu$ and 
is capable of specifying bounded inevitabilities.  

Our tool {\tt red} (version 4.1\cite{Wang03}) is a full TCTL model-checker/simulator 
with a BDD-like data-structure, 
called CRD (clock-restriction diagram)\cite{Wang01a,Wang01b,Wang03}.  
Previous research with {\tt red} has focused on enhancing the performance of 
safety analysis.  

Abstraction techniques for analysis have been studied in great depth
since the pioneering work of Cousot et al\cite{CC77,CC92a}.  
For TA, convex-hull over-approximation\cite{WongToi95} has been a popular choice
for DBM technology due to its intuitiveness and effective performance.  
It is difficult to implement this over-approximation in {\tt red}\cite{Wang03} since
variable-accessing has to observe variable-orderings of BDD-like data-structures.  
Nevertheless, many over-approximation techniques for TA have been reported in 
\cite{Balarin96} for BDD-like data-structures and in \cite{WHY03} specifically for CRD.  

Relations between abstraction techniques and subclasses of CTL with only universal 
(or existential respectively) path quantifiers has been studied in \cite{CGJLV00,LPJHS96}.  
As mentioned above, the corresponding framework in TCTL is noted in \cite{Moller02}.

\section{Timed automata (TA) \label{sec.system}}

We use the widely accepted model of {\em timed automata}\cite{AD89}, which
is a finite-state automata equipped with 
a finite set of clocks which can hold nonnegative real-values.  
At any moment, a timed automata can stay in only one {\em mode} (or {\em control location}).  
In its operation, one of the transitions can be triggered
when the corresponding triggering condition is satisfied.  
Upon being triggered, the automata instantaneously transits from one
mode to another and resets some clocks to zero. 
Between transitions, all clocks increase readings at a uniform 
rate.  

For convenience, given a set $Q$ of modes and a set $X$ of clocks, we use $B(Q,X)$ as
the set of all Boolean combinations of atoms of the forms $q$ and 
$x-x'\sim c$,  where $q\in Q$, $x,x'\in X\cup\{0\}$, ``$\sim$'' is one of 
$\leq, <,=,>,\geq$, and $c$ is an integer constant.
{\definition \underline{\bf timed automata (TA)}}
A TA $A$ is given as 
a tuple $\langle X, Q, I, \mu, T, \tau, \pi\rangle$ 
with the following restrictions. 
$X$ is the set of clocks.  
$Q$ is the set of  modes.
$I\in B(Q,X)$ is the initial condition.  
$\mu:Q\mapsto B(\emptyset,X)$ defines the invariance condition of each mode.
$T\subseteq Q\times Q$ is the set of transitions.  
$\tau:T\mapsto B(\emptyset,X)$ and $\pi:T\mapsto 2^X$ respectively
define the triggering condition and the clock set to reset 
of each transition.  
\qed 
\vspace*{2mm} 

A {\em valuation} of a set is a mapping from the set to another set.  
Given an $\eta\in B(Q,X)$ and a valuation $\nu$ of $X$, we say 
$\nu$ {\em satisfies} $\eta$, in symbols $\nu\models\eta$,
iff it is the case that 
when the variables in $\eta$ are interpreted according to $\nu$, 
$\eta$ will be evaluated as $\true$.  
{\definition \underline{\bf states}}  
A state $\nu$ of $A=\langle X, Q, I, \mu, T, \tau, \pi\rangle$ is a valuation of
$X\cup Q$ such that 
\begin{list1}
\item there is a unique $q\in Q$ such that $\nu(q)=\true$ and 
	for all $q'\neq q, \nu(q')=\false$; 
\item for each $x\in X$, $\nu(x)\in {\cal R}^+$ 
	(the set of nonnegative reals) 
	and $\forall q\in Q, \nu(q)\Rightarrow \nu\models\mu(q)$.
\end{list1}
Given state $\nu$ and $q\in Q$ such that $\nu(q)=\true$, 
we call $q$ the mode of $\nu$, in symbols $\nu^Q$.  
\qed 
For any $t\in {\cal R}^+$, $\nu+t$ is a state identical to $\nu$ 
except that for every clock $x\in X$, $\nu(x)+t = (\nu+t)(x)$.
Given $\bar{X}\subseteq X$, $\nu \bar{X}$ is a new state identical 
to $\nu$ except that for every $x\in \bar{X}$, $\nu\bar{X}(x)=0$.

{\definition \underline{\bf runs (computations)}}  
Given a TA $A=\langle X, Q, I, \mu, T, \tau, \pi\rangle$,   
a {\em run} is an (infinite) sequence of state-time pairs, 
$(\nu_0,t_0)(\nu_1,t_1)\ldots(\nu_k,t_k)\ldots\ldots$
, such that $\nu_0\models I$ and $t_0 t_1 \ldots t_k\ldots\ldots$
is a monotonically increasing real-number (time) 
divergent sequence, 
and for all $k\geq 0$, 
\begin{list1}
\item {\em invariance conditions are preserved in each interval:} that is,\\
	for all $t\in [0, t_{k+1}- t_k]$,
	$\nu_k+t\models\mu(\nu_k^Q)$; and 
\item either {\em no transition happens at time $t_k$}, that is, 
	$\nu_k^Q=\nu_{k+1}^Q$ 
	and $\nu_k+(t_{k+1}-t_k) = \nu_{k+1}$; 
	or 
	{\em a transition happens at $t_k$}, that is, 
	\begin{list2}
	\item {\em there is such a transition}, that is 
		$(\nu_k^Q,\nu_{k+1}^Q)\in T$; and
	\item {\em the corresponding triggering condition is satisfied}, that is, 
		$\nu_k+(t_{k+1}-t_k)\models\tau(\nu_k^Q,\nu_{k+1}^Q)$; and 
	\item {\em the clocks are reset to zero accordingly}, 
		that is, $(\nu_k+(t_{k+1}-t_k))
		\pi(\nu_k^Q,\nu_{k+1}^Q)
		=\nu_{k+1}$.
\qed
 	\end{list2}
\end{list1}

\section{TCTL (Timed CTL) \label{sec.tctl}}

TCTL\cite{ACD90,HNSY92} is a branching temporal logic
for the specification of dense-time systems.
{\definition\label{def.syntax.tctl}{\bf (Syntax of TCTL formulas):}}
A TCTL formula $\phi$ has the following syntax rules.
\begin{center}
$\begin{array}{rrl}
\phi & ::= & \eta\,|\, \neg\phi_1  \,|\, \phi_1\vee \phi_2 \,|\,
        x.\phi_1 \,|\,
        \exists \phi_1 \until \phi_2 \,|\,
        \exists \pfrr \phi_1 
\end{array}$
\end{center}
Here $\eta\in B(Q,X)$ and $\phi_1$, $\phi_2$ are TCTL formulas.
\qed 

The modal operators are intuitively explained in the following.
\begin{list1}
\item $x.\phi$ means that "if there is a clock $x$ with reading zero now,
	then $\phi$ is satisfied."
\item $\exists$ means ``there exists a run''
\item $\phi_1\until\phi_2$
	means that along a computation, $\phi_1$ is true until
	$\phi_2$ becomes true.
\item $\pfrr\phi_1$
	means that along a computation,
	$\phi_1$ is always true.
\end{list1}
Besides the standard shorthands of temporal logics\cite{ACD90,HNSY92},
we adopt the following for TCTL:
\begin{center}
$\begin{array}{llll} 
\bullet		& \exists\pevt \phi_1\mbox{ for }\exists\true\; \until \phi_1
& \bullet	& \forall\pfrr \phi_1\mbox{ for }\neg\exists\pevt \neg \phi_1 \\
\bullet		& \forall\phi_1\until \phi_2 \mbox{ for } 
    			\neg((\exists(\neg\phi_2)\until
                		\neg(\phi_1\vee\phi_2))\vee
           			(\exists\pfrr\neg\phi_2))
& \bullet	& \forall\pevt \phi_1\mbox{ for }\forall\true\until\phi_1
\end{array}$
\end{center}

{\definition\label{def.satisfy.tctl}{\bf (Satisfaction of TCTL formulas):}}
We write in notations $A,\nu\models \phi$ to mean that
$\phi$ is satisfied at state $\nu$ in TA $A$.
The satisfaction relation is defined inductively as follows.
\begin{list1}
\item When $\phi_1\in B(Q,X)$,
	$A,\nu\models \phi_1$ according to the definition in the beginning of
	subsection~\ref{sec.system}.
\item $A,\nu\models \phi_1 \vee \phi_2$ iff either $A,\nu\models \phi_1$ or $A,\nu\models \phi_2$
\item $A,\nu\models \neg \phi_1$ iff $A,\nu\not\models \phi_1$
\item $A,\nu\models x.\phi_1$ iff $A,\nu\{x\}\models \phi_1$.
\item $A,\nu\models \exists\phi_1\until\phi_2$
        iff there exists a run
	$(\nu_1,t_1)(\nu_2,t_2)\ldots$ such that
	$\nu_1=\nu$ in $A$, and there exist
	an $i\geq 1$ and a $\delta\in [0,t_{i+1}-t_i]$, s.t.
        \begin{list2}
        \item $A,\nu_i+\delta\models \phi_2$,
        \item for all $j, \delta'$, if either
        	$(1\leq j<i)\wedge(\delta'\in[0,t_{j+1}-t_j])$ or
        	$(j=i)\wedge (\delta'\in[0,\delta))$, then
                $A, \nu_j+\delta'\models \phi_1$.
        \end{list2}
        In words, $\nu$ satisfies
        	$\exists\phi_1\until\phi_2$
        iff there exists a run from $\nu$ such that
        along the run, $\phi_1$ is true until $\phi_2$ is true.
\item $A,\nu\models \exists\pfrr\phi_1$
        iff there exists a run
        $(\nu_1,t_1)(\nu_2,t_2)\ldots$ such that $\nu_1=\nu$ in $A$,
        and for every $i\geq 1$ and $\delta\in [0,t_{i+1}-t_i]$,
        $A,\nu_i+\delta\models\phi_1$.
        In other words, $\nu$ satisfies $\exists\pfrr\phi_1$
        iff there exists a run from $\nu$ such that $\phi_1$ is always true.
\end{list1}
A TA $A$ satisfies
a TCTL formula $\phi$, in symbols $A\models \phi$, iff
for every state $\nu_0\models I$,
$A,\nu_0\models \phi$.
\qed

\section{TCTL Model-checking algorithm with non-Zeno requirements\label{sec.algorithm}}

Our model-checking algorithm is backward reasoning.  
We need two basic procedures, one for the computation of the weakest precondition of
transitions, and the other for backward time-progression.
These two procedures are important in the symbolic construction of backward
reachable state-space representations.
Various presentations of the two procedures can be found in
\cite{HNSY92,Wang00a,Wang00b,Wang01a,Wang01b,Wang03,WH02}.
Given a state-space representation $\eta$ and a transition $e$,
the first procedure, $\mbox{\tt xtion\_bck}(\eta, e)$, computes the weakest precondition
\begin{list1}
\item in which, every state satisfies the invariance condition imposed by $\mu()$; and
\item from which we can transit to states in $\eta$ through $e$.
\end{list1}
The second procedure, $\mbox{\tt time\_bck}(\eta)$, computes the space representation of states
\begin{list1}
\item from which we can go to states in $\eta$ simply by time-passage; and
\item every state in the time-passage also satisfies the invariance condition imposed by
	$\mu()$.
\end{list1}
With the two basic procedures, we can construct a symbolic backward reachability procedure
as in \cite{HNSY92,Wang00a,Wang00b,Wang01a,Wang01b,Wang03,WH02}.
We call this procedure $\breach(\eta_1,\eta_2)$ for convenience.
Intuitively, $\breach(\eta_1,\eta_2)$ characterizes the
backwardly reachable state-space from states in $\eta_2$ through runs along
which all states satisfy $\eta_1$.
Computationally, $\breach(\eta_1,\eta_2)$ can be defined as
the least fixpoint of the equation
$Y=\eta_2\vee \left(\eta_1\wedge\mbox{\tt time\_bck}
	(\eta_1\wedge\bigvee_{e\in T}\mbox{\tt xtion\_bck}(Y, e))\right)$, 
i.e.,
\begin{center}
$\breach(\eta_1,\eta_2)\equiv \mbox{\tt lfp } 
	Y.\left(\eta_2\vee \left(\eta_1\wedge\mbox{\tt time\_bck}
	(\eta_1\wedge\bigvee_{e\in T}\mbox{\tt xtion\_bck}(Y, e))\right)\right)$.
\end{center}

Our model-checking algorithm is modified from the classic model-checking algorithm
for TCTL\cite{HNSY92}.
The design of the greatest fixpoint evaluation algorithm with consideration of 
non-Zeno requirement is based on the following lemma. 
{\lemma \label{lemma.gfp} Given $d\geq 1$, $A,\nu\models\exists\pfrr\eta$ iff
there is a finite run from $\nu$ of duration $\geq d$ such that
along the run every state satisfies $\eta$ and
the finite run ends at a state satisfying $\exists\pfrr\eta$.
}
\\\pf Details are omitted due to page-limit. 
But note that we can construct an infinite and divergent run 
by concatenating an infinite sequence of finite runs with durations $d\geq 1$.
The existence of infinitely many such concatenable finite runs is assured by  
the recursive construction of $\exists\pfrr\eta$.  
\qed

Then $\exists\pfrr\eta$ can be defined with the following greatest fixpoint. 
\begin{center} 
$\exists\pfrr\eta\equiv 
\mbox{\tt gfp } Y. \left(\mbox{\tt ZC}.\;\breach(\eta,Y\wedge\mbox{\tt ZC}\geq d)\right)$
\end{center} 
Here clock {\tt ZC} is used specifically to measure the non-Zeno requirement.  
The following procedure can construct the greatest fixpoint 
satisfying $\exists\pfrr\eta$ with a non-Zeno requirement.
\hrule
\vspace*{2mm}
\noindent
$\mbox{\tt gfp}(\eta)$ /* d is a static parameter for measuring time-progress */ \{
\hst	$Y:=\eta$; $Y':=\true$; 
\hst	repeat until $Y= Y'$, \{\hfill (1)
\hstt		$Y':=Y$; 
		$Y:=Y\wedge\mbox{\tt clock\_eliminate}(\mbox{\tt ZC}=0
				\wedge\breach(\eta,Y\wedge \mbox{\tt ZC}\geq d),
					\mbox{\tt ZC})$; \hfill (2)
\hst	\}
\hst	return $Y$;
\\\}
\vspace*{2mm}
\hrule
\vspace*{3mm}
Here $\mbox{\tt clock\_eliminate}()$ removes a clock from a state-predicate without
losing information on relations among other clocks.
Details can be found in appendix~\ref{app.clock.eliminate}.  

Note here that $d$ works as a parameter.
We can choose the value of $d\geq 1$ for better performance in the computation
of the greatest fixpoint.

Procedure {\tt gfp}() can be used in the labeling algorithm in \cite{ACD90,HNSY92}
to replace the evaluation of $\exists\pfrr$-formulas. 
For completeness of the presentation, please check appendix~\ref{app.mck} 
to see our complete model-checking algorithm with non-Zeno requirement.  
The correctness follows from Lemma~\ref{lemma.gfp}.

\section{Model-checking with Early Decision on the Greatest Fixpoint (EDGF) \label{sec.edgf}}

In the evaluation of the greatest fixpoint for formulas like $\exists\pfrr\phi_1$,
we start from the description, say $Y$, for a subspace of $\phi_1$ and iteratively eliminate those
subspaces which cannot go to a state in $Y$ through finite runs of $d\geq 1$ time units.
Thus, the state-space represented by $Y$ shrinks iteratively until it settles at a
fixpoint.
In practice, this greatest fixpoint usually happens in conjunction with other formulas.
For example, we may want to specify
$\mbox{\tt collision}\rightarrow y.\forall \pevt (y<26\wedge \mbox{\tt idle})$
meaning that a bus at the collision state, will enter the idle state in 26 time-units.
After negation for model-checking,
we get
$\mbox{\tt collision}\wedge y.\exists\pfrr (y\geq 26\vee \neg\mbox{\tt idle})$.  
In evaluating this negated formula, we want to see if the greatest fixpoint for
the $\exists\pfrr$-formula intersects with the state-space for $\mbox{\tt collision}$.
We do not actually have to compute the greatest fixpoint to know if the intersection is
empty.
Since the value of $Y$ iteratively shrinks, we can check if the intersection between $Y$
and the state-space for $\mbox{\tt collision}$ becomes empty at each iteration of
the greatest fixpoint construction (i.e., the repeat-loop at statement (1) in procedure
{\tt gfp}).
If at an iteration, we find the intersection with $Y$ is already empty, then there is no
need to continue calculating the greatest fixpoint and we can immediately return the current value of $Y$
(or $\false$) without affecting the result of the model-checking.

Based on this idea, we rewrite our model-checking algorithm with our {\em Early Deision on
the Greatest Fixpoint (EDGF)}.
We introduce a new parameter $\beta$ to pass the information of
the target states inherited from the scope.
\vspace*{3mm}
\hrule
\vspace*{2mm}
\noindent $\evalg(A,\chi,\beta,\bar{\phi})$ \\
/* $\chi$ is the set of clocks declared in the scope of $\bar{\phi}$ */ \\
/* $\beta$ is constraints inhereted in the scope of $\bar{\phi}$ for early decision of gfp */ \{
\hst switch ($\bar{\phi}$) \{
\hst {\bf case} ($\false$):
	return $\false$;
\hst {\bf case}  ($p$):
	return $p\wedge\bigwedge_{x\in \chi}x\geq 0$;
\hst {\bf case}  ($x-y\sim c$):
	return $x-y\sim c\wedge_{x\in \chi}x\geq 0$;
\hst {\bf case}  ($\phi_1\vee \phi_2$):
	return $\evalg(A,\chi,\beta,\phi_1)\vee \evalg(A,\chi,\beta,\phi_2)$;
\hst {\bf case}  ($\phi_1\wedge \phi_2$):
\hstt 	if $\phi_2$ does not contain modal operator, \{
\hsttt 		$\eta_2:=\evalg(A,\chi,\beta,\phi_2)$;
\hsttt		return $\eta_2\wedge \evalg(A,\chi,\beta\wedge\eta_2,\phi_1)$;\hfill (3)
\hstt	\}
\hstt	else \{
\hsttt		$\eta_1:=\evalg(A,\chi,\beta,\phi_1)$;
\hsttt		return $\eta_1\wedge \evalg(A,\chi,\beta\wedge\eta_1,\phi_2)$;\hfill (4)
\hstt	\}
\hst {\bf case}  ($\neg \phi_1$):
	return $\neg \evalg(A,\chi,\true,\phi_1)$;
\hst {\bf case}  ($x.\phi_1$):
	return $\mbox{\tt clock\_eliminate}(x=0\wedge \evalg(A,\chi\cup\{x\},\beta,\phi_1\wedge x\geq 0), x)$;
\hst {\bf case}  ($\exists \phi_1\until\phi_2$):
\hstt	$\eta_1:= \evalg(A,\chi,\true,\phi_1)$;
 	$\eta_2:= \evalg(A,\chi,\true,\phi_2)$;
\hstt	return $\breach(\eta_1,\eta_2)$;
\hst {\bf case}  ($\exists\pfrr\phi_1$): 
	return $\mbox{\tt gfp\_EDGF}(\evalg(\phi_1),\beta)$;
\hst\}
\\\}
\vspace*{2mm}
\\
$\mbox{\tt gfp\_EDGF}(\eta,\beta)$ /* d is a static parameter for measuring time-progress */ \{
\hst	$Y:=\eta$; $Y':=\true$; 
\hst	repeat until $Y= Y'$ or $(Y\wedge \beta)=\false$, \{\hfill (5)
\hstt		$Y':=Y$; 
		$Y:=Y\wedge\mbox{\tt clock\_eliminate}(\mbox{\tt ZC}=0
				\wedge\breach(\eta,Y\wedge \mbox{\tt ZC}\geq d),
					\mbox{\tt ZC})$; 
							\hfill (6) 
\hst	\}
\hst	return $Y$;
\\\}
\vspace*{2mm}
\hrule
\vspace*{3mm}
To model-check TA $A$ against TCTL formula $\phi$, 
we reply $\true$ iff $\evalg(A,\emptyset,\true,\neg\phi)$ is false.  
As can be seen from statement (3) and (4) in the case of conjunction formulas,
we strengthen the target-state information.
In the evaluation of the greatest fixpoint, we use condition-testing
$(Y\wedge \beta)=\false$
in statement (5) respectively to check for early decision.

\section{Greatest fixpoint computation by tolerating Zenoness
	\label{sec.Zeno}
	}

In practice, the greatest fixpoint computation procedures presented in the last
two sections can be costly in computing resources since their characterizations have
a least fixpoint nested in a greatest fixpoint.
This is necessary to guarantee that only nonZeno computations are considered.
In reality, it may happen that, due to well-designed behaviors,
systems may still satisfy certain inevitability properties for both Zeno and non-Zeno
computation.
In this case, we can benefit from a less expensive procedure to compute the greatest fixpoint.
For example, we have designed the following procedure which does not rule out
Zeno computations in the evaluation of $\exists\pfrr$-formulas.
\vspace*{3mm}
\hrule
\vspace*{2mm}
\noindent
$\mbox{\tt gfp\_Zeno\_EDGF}(\eta,\beta)$ \{
\hst	$Y:=\eta; Y':=\true$;
\hst	repeat until $Y= Y'$ or $(Y\wedge \beta)=\false$, \{ \hfill (7)
\hstt		$Y':=Y$; 
		$Y:=Y\wedge\eta\wedge\mbox{\tt time\_bck}
		\left(\eta\wedge\bigvee_{e\in T}\mbox{\tt xtion\_bck}(Y, e)\right)$;
\hst	\}
\hst	return $Y$;
\\\}
\vspace*{2mm}
\hrule
\vspace*{3mm}
Even if the procedure can be imprecise in over-estimation
of the greatest fixpoint,
it can be much less expensive in the verification of well-designed
real-world projects.

\section{Abstract model-checking with \tctla \label{sec.tctle}}

We have also experimented with abstraction techniques in the evaluation of 
greatest fixpoints. 
Due to page-limit, we shall leave the explanation in appendix~\ref{app.tctla}.  
The corresponding xperiment report is in subsection~\ref{subsec.exp.tctla}.

\section{Implementation and experiments\label{sec.experiments}}

We have implemented the ideas in our model-checker/simulator, {\tt red} version 4.1, for TA.
{\tt red} uses the new BDD-like data-structure,
{\em CRD} (Clock-Restriction Diagram)\cite{Wang01a,Wang01b},
and supports both forward and backward analysis,
full TCTL model-checking with non-Zeno computations,
deadlock detection, and counter-example generation.
Users can also declare global and local (to each process)
variables of type clock, integer, and pointer (to identifier of processes).
Boolean conditions on variables can be tested and
variable values can be assigned.
The TCTL formulas in {\tt red} also allow quantification on process identifiers for
succinct specification.
Interested readers can download {\tt red} for free from
\begin{verbatim}
                             http://cc.ee.ntu.edu.tw/~val/
\end{verbatim}
We design our experiment in two ways. 
First, we run {\tt red} 4.1 with various options and benchmarks to test
if our ideas can indeed improve the verification performance of inevitability properties 
in $\tctla$.  
Second, we compare {\tt red} 4.1 with Kronos 5.2 to check if our implementation remains competitive
in regard to other tools.  
However, we remind the readers that comparison report with other tools should be 
read carefully since {\tt red} uses different data-structures from Kronos.  
Moreover, it is difficult to know what fine-tuning techniques each tool has used. 
Thus it is difficult to conclude if the techniques presented in this work 
really contribute to the performance difference between {\tt red} and Kronos.  
Nevertheless, we believe it is still an objective measure to roughly estimate 
how our ideas perform.  

In the following section, we shall first discuss the design of our benchmarks, 
then report our experiments.  
Data is collected on a Pentium 4 1.7GHz with 256MB memory running LINUX.
Execution times are collected for Kronos while
times and memory (for data-structure) are collected for {\tt red}.
"s" means seconds of CPU time, 
"k" means kilobytes for memory space for data-structures,
"O/M" means "out-of-memory."

\subsection{Benchmarks \label{subsec.benchmarks}} 

We do not claim that the benchmarks selected here represent the complete spectrum of 
model-checking tasks.  
The evaluation of TCTL formulas may incur various complex computations 
depending on the structures of the timed automata and the specification formulas.  
But we do carefully choose our benchmarks according to the broad spectrum of combination 
of models and specifications so that we can gain some insights about performance 
enhancement of TCTL inevitability analysis.  
Benchmarks include three different timed automatas and specifications 
for unbounded inevitability, bounded inevitability\cite{HNSY92}, and
modal operators with nesting depth zero, one, and two respectively.  
We identify one important benchmark which can only be verified 
with non-Zeno computations.  
The other benchmarks can be (safely) verified 
without requirement of non-Zeno computations.  

Due to page-limit, we leave the description of the benchmarks in appendix~\ref{app.benchmarks}.

\subsection{Performance w.r.t. parameter for measuring time-progress \label{subsec.d}}

In statement (2) of procedure {\tt gfp}() and 
statement (6) of procedure {\tt gfp\_EDGF}(), 
we use inequality $\mbox{\tt ZC}\geq d$ to check 
time-progress in non-Zeno computations, where $d$ is a parameter $\geq 1$.  
We can choose various values for the parameter in our implementations.  
In our experiment reported in this subsection, we have found that  
the value of parameter $d$ can greatly affect the verification performance.  

In this experiment, we shall use various values of parameter $d$ ranging from 
$1$ to beyond the biggest timing constants used in the models.  
For the leader-election benchmark, the biggest timing constant used is $2$.  
For the Pathos benchmark, the biggest timing constant used is equal to the number of 
processes.  
For the CSMA/CD benchmarks (A), (B), and (C), the 
biggest timing constant used is equal to 808.  

\begin{figure}
\begin{center}
\begin{tabular}{cc} 
\epsfig{width=70mm, file=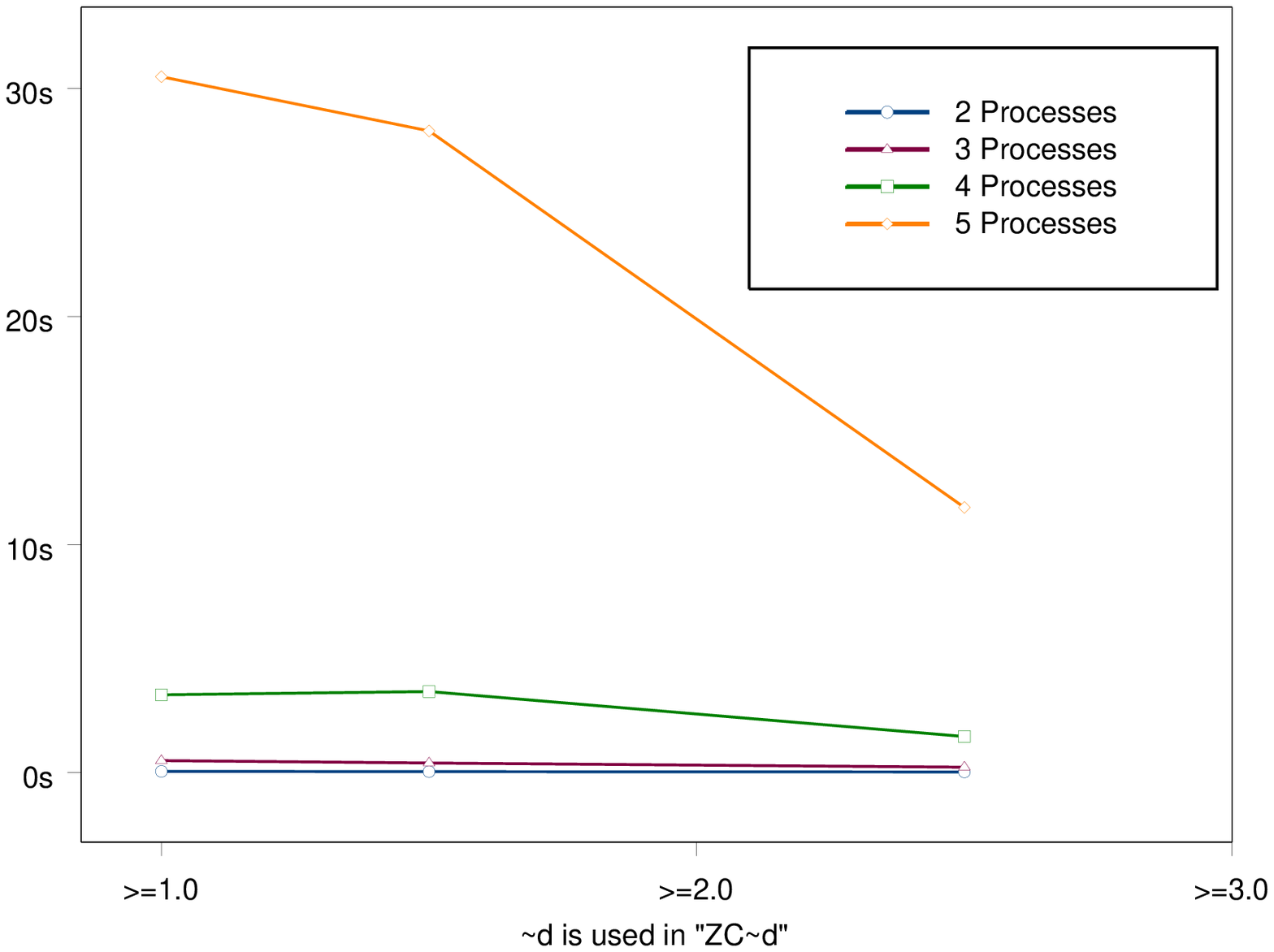} 
& \epsfig{width=70mm, file=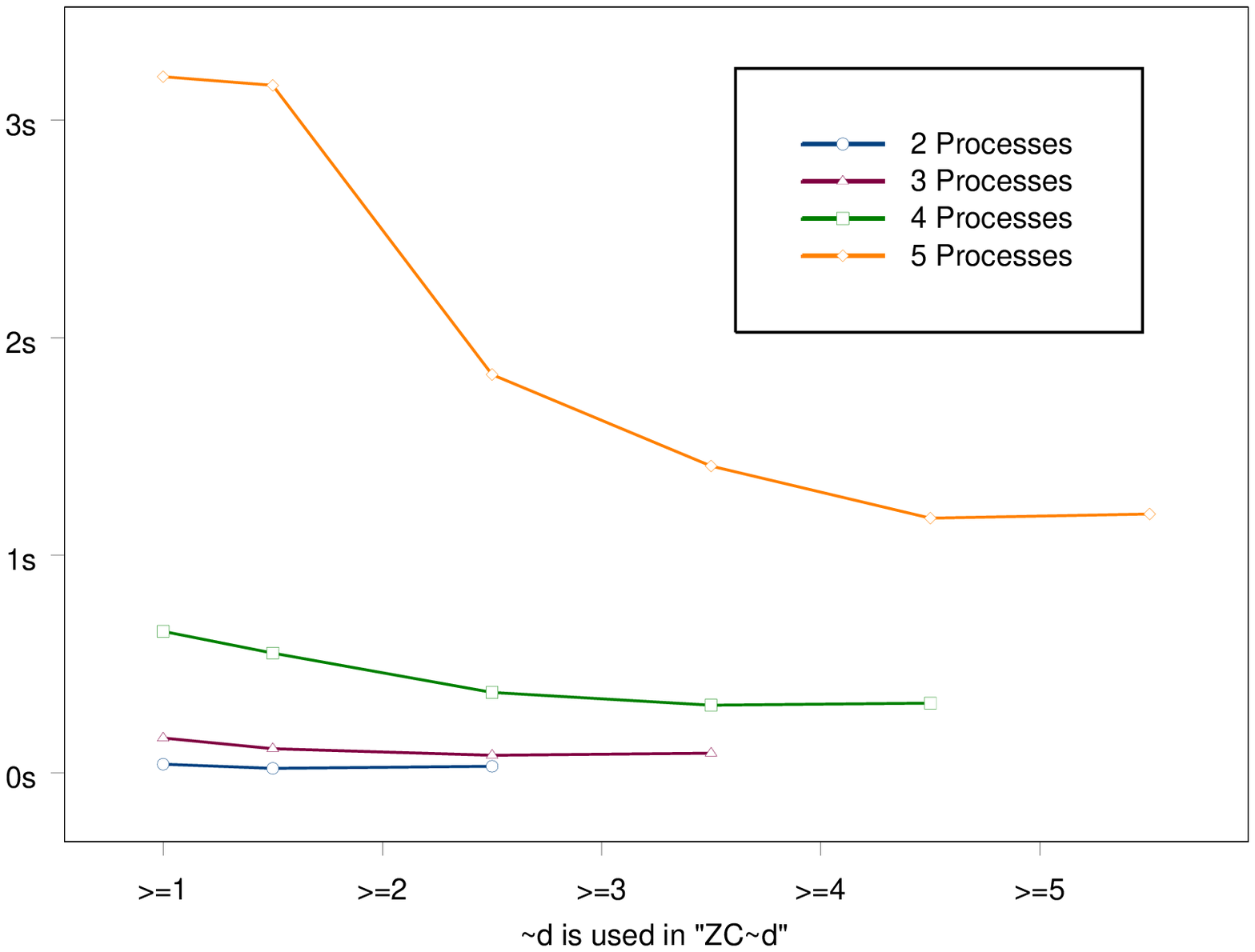} \\
(a) leader-election & (b) PATHOS \\ 
\epsfig{width=70mm, file=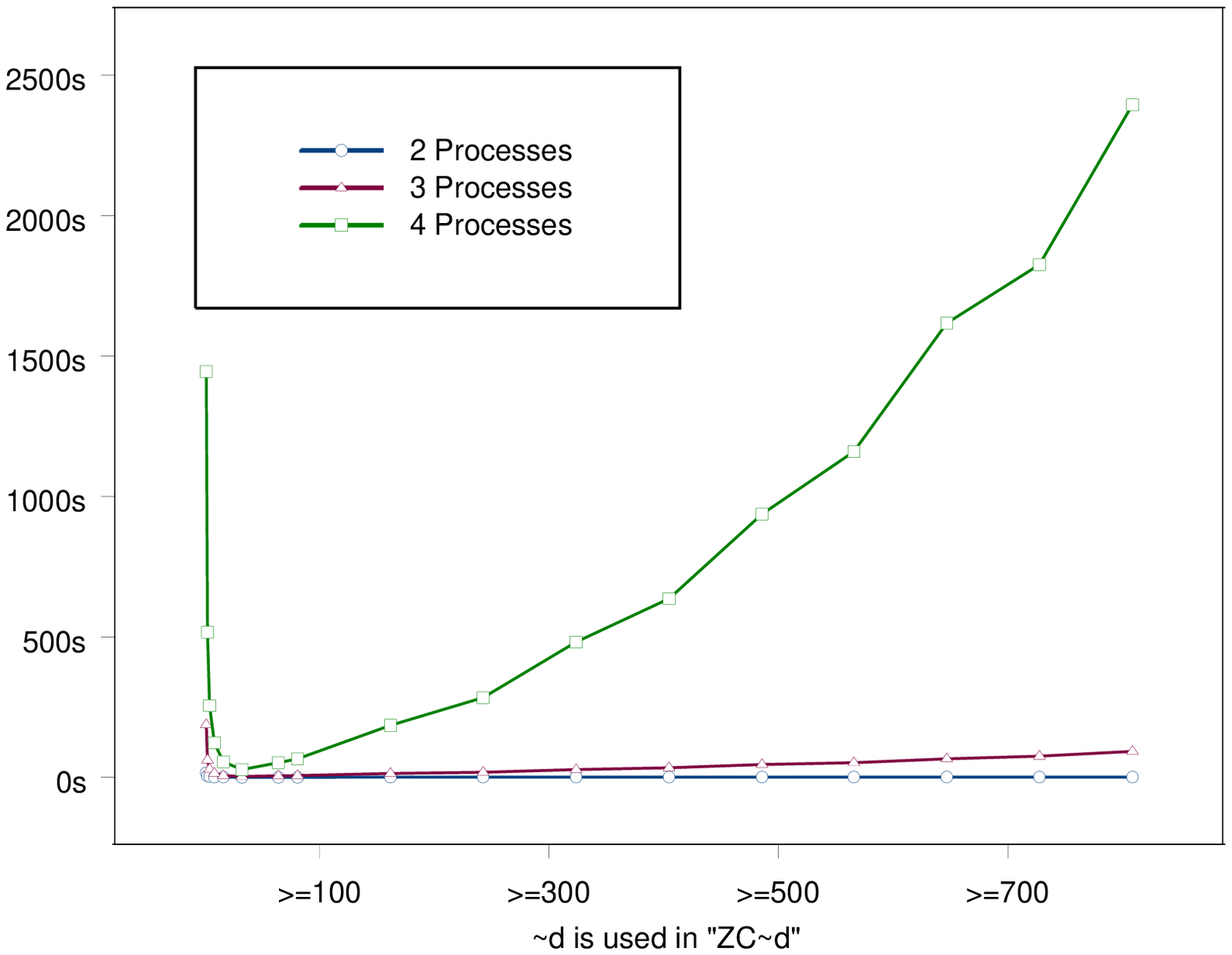}
& \epsfig{width=70mm, file=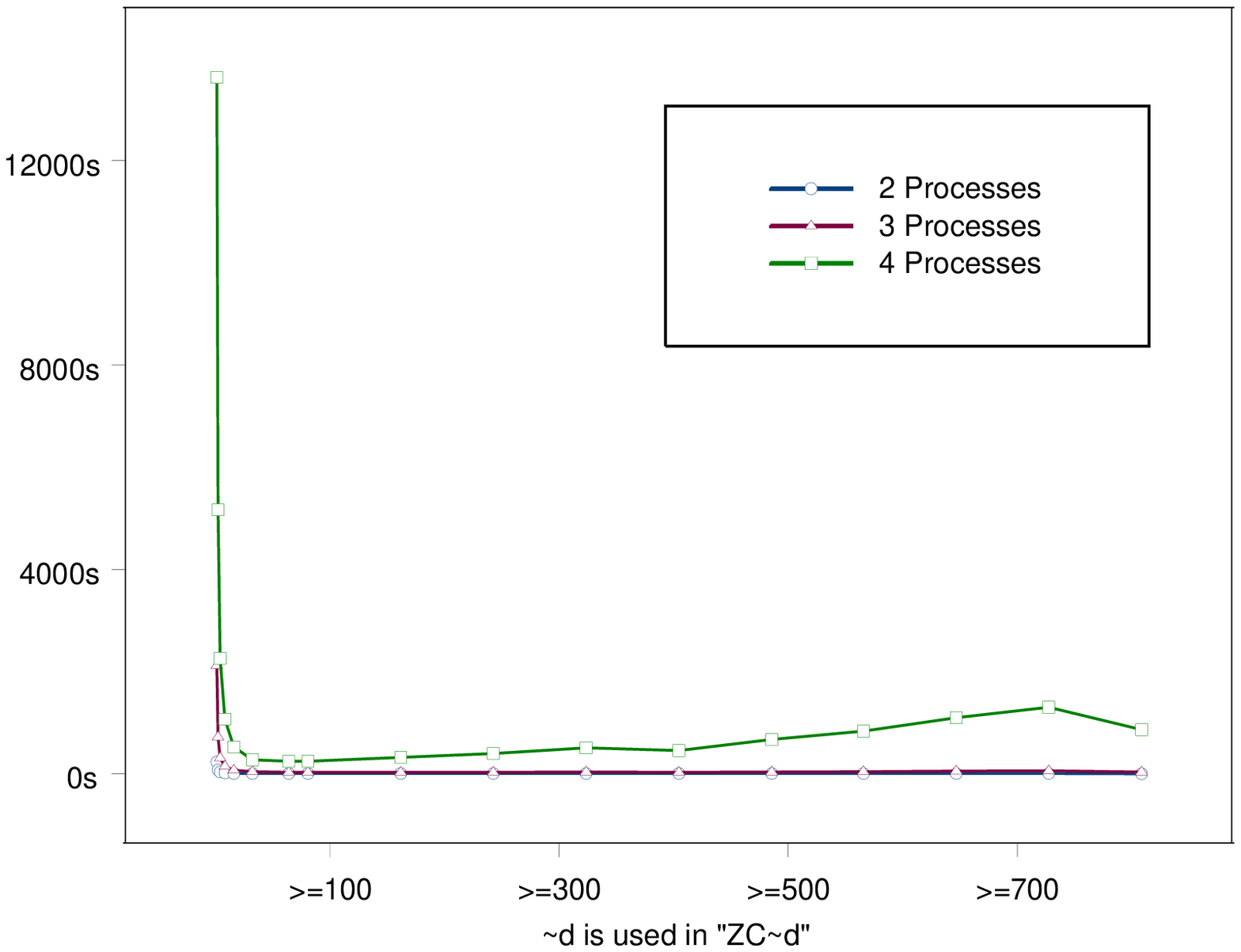}\\
(c) CSMA/CD(A) & (d) CSMA/CD(B) \\
\multicolumn{2}{c}{\epsfig{width=70mm, file=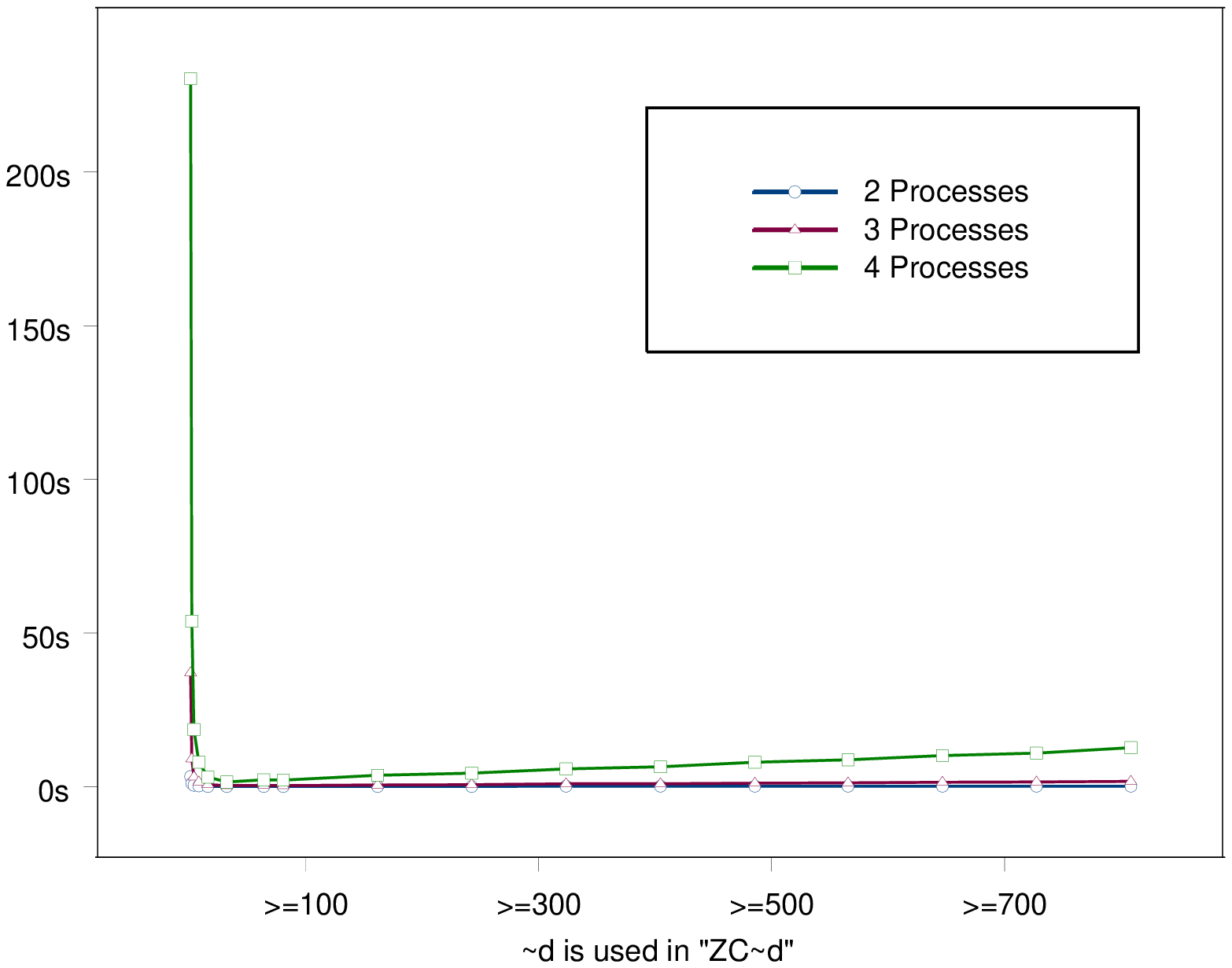}}\\
\multicolumn{2}{c}{(e) CSMA/CD(C)}
\end{tabular}
\end{center}
The Y-axis is with "time in sec"   
while the X-axis is with "$\sim d$" used in "$\mbox{\tt ZC}\sim d$."  
\caption{Time-complexity charts w.r.t. $d$-values (Data collected with option EDGF)} 
\label{fig.charts.time} 
\end{figure}

In fact, we can also use inequality $\mbox{\tt ZC}>d$, with $d\geq 1$, 
in statements (2) and (6) 
of procedures {\tt gfp}() and {\tt gfp\_EDGF}() respectively.  
Due to page-limit, we shall leave the performance data table to 
appendix~\ref{app.d.table}.  
We have drawn charts to show time-complexity for the benchmarks w.r.t. $d$-values
in figure~\ref{fig.charts.time}.  
More charts for the space-complexity can be found in appendix~\ref{app.d.mem}.  

As can be seen from the charts, our algorithms may respond with different complexity 
curves to various model structures and specifications.  
For benchmarks leader-election and PATHOS, 
it seems that the bigger the $d$-value, the better the performance.  
For the three CSMA/CD benchmarks, it seems that 
the best performance happens when $d$ is around 80.  
But one thing common in these charts is that $d=1$ always gives the worst performance.  

We have to admit that we do not have a theory to analyze or predict 
the complexity curves w.r.t. various model structures and specifications.  
More experiments on more benchmarks may be needed in order to get more understanding 
of the curves.  
In general, we feel it can be difficult to analyze such complexity curves.   
After all, our models of TA are still "programs" in some sense.  

Nevertheless, we have still tried hard to look into the execution of our algorithms 
for explanation of the complexity cuves.  
Procedures {\tt gfp}() and {\tt gfp\_EDGF}() both are constructed with an 
inner loop (for the least fixpoint evaluation of {\tt reachable-bck}()) and 
an outer loop (for the greatest fixpoint evaluation).  
With bigger $d$-values, 
it seems that the outer loop converges faster 
while the inner loop converges slower.  
That is to say, with bigger $d$-values, we may need less iterations of the outer-loop 
and, in the same time, more iterations of the inner loop to compute
the greatest fixpoints.  
The complexity patterns in the charts are thus superpositions between 
the complexities of the outer loop and the inner loop. 

We have used the $d$-values with the best performance for 
the experiments reported in the next few subsections.  
For benchmarks PATHOS and leader-election, $\sim d$ is set to $>C_{A:\phi}$.   
($C_{A:\phi}$ is the biggest timing constant used in model $A$ and TCTL specification $\phi$.)  
For the three CSMA/CD benchmarks, $\sim d$ is set to $\geq 80$.

\subsection{Performance w.r.t. non-Zeno requirement and EDGF}

In our first experiment, we observe the performance of
our inevitability analysis algorithm w.r.t. the non-Zeno requirement and
the EDGF policy.
The performance data is in table~\ref{tab.Z}.
\begin{table}
\begin{center}
\scriptsize 
\begin{tabular}{|l|l||r|r|r|r|}
\hline
& & \multicolumn{2}{c|}{no non-Zeno requirement} & \multicolumn{2}{c|}{non-Zeno requirement}  \\ \cline{3-6}
benchmarks  &concurrency& \multicolumn{1}{c|}{EDGF}
			& \multicolumn{1}{c|}{no EDGF}
			& \multicolumn{1}{c|}{EDGF}
			& \multicolumn{1}{c|}{no EDGF}	\\\cline{3-6}
	&		& time/space/answer	& time/space/answer	& time/space/answer	& time/space/answer \\ \hline
pathos	& 2 proc.s	& 0.02s/7k/true		& 0.02s/7k/true		& 0.03s/7k/true		& 0.03s/7k/true	\\ \cline{2-6}
	& 3 proc.s	& 0.09s/18k/true	& 0.1s/18k/true		& 0.08s/17k/true	& 0.09s/17k/true \\ \cline{2-6}
	& 4 proc.s	& 0.63s/74k/true	& 0.66s/74k/true	& 0.31s/42k/true   	& 0.31s/42k/true \\ \cline{2-6}
	& 5 proc.s	& 6.52s/857k/true	& 6.65s/859k/true	& 1.17s/114k/true  	& 1.28s/114k/true \\ \cline{2-6}
	& 6 proc.s	& 161s/15087k/true	& 162s/15090k/true	& 5.22s/314k/true	& 5.37s/314k/true \\ \cline{2-6}
	& 7 proc.s	& O/M			& O/M			& 30.71s/942k/true	& 31.16s/941k/true \\ \hline
leader	& 2 proc.s	& 0.04s/10k/true	& 0.03s/10k/true	& 0.04s/16k/true	& 0.04s/16k/true  \\ \cline{2-6}
election& 3 proc.s	& 0.28s/33k/true 	& 0.28s/33k/true	& 0.25s/84k/true	& 0.24s/84k/true \\ \cline{2-6}
	& 4 proc.s	& 1.96s/84k/true	& 1.98s/84k/true	& 1.54s/338k/true	& 1.53s/338k/true \\ \cline{2-6}
	& 5 proc.s	& 10.01s/23/true	& 10.07s/234k/true	& 11.23s/1164k/true 	& 11.17s/1164k/true \\ \cline{2-6}
	& 6 proc.s	& 52.63s/635k/true	& 48.28s/635k/true	& 110.9s/7992k/true	& 110.2s/7992k/true \\ \cline{2-6}
	& 7 proc.s	& 206.7s/1693k/true	& 205.7s/1693k/true	& 860.5s/42062k/true	& 859.7s/42062k/true\\ \hline
CSMA/CD	& bus+2 senders	& 0.07s/25k/true	& 0.15s/25k/true	& 0.33s/42k/true 	& 9.29s/90k/true\\ \cline{2-6}
(A)	& bus+3 senders	& 0.24s/49k/true 	& 0.66s/63k/true	& 3.09s/191k/true	& 98.33s/191k/true\\ \cline{2-6}
	& bus+4 senders	& 0.78s/131k/true 	& 2.38s/201k/true	& 26.23s/936k/true	& 867.5s/1578k/true\\ \cline{2-6}
	& bus+5 senders	& 2.39s/378k/true	& 8.47s/625k/true 	& 195.14s/4501k/true 	& 6021s/7036k \\ \hline
CSMA/CD	& bus+2 senders	& 0.16s/25k/maybe	& 0.16s/25k/maybe	& 1.92s/37k/true	& 2.3s/37k/true \\ \cline{2-6}
(B)	& bus+3 senders	& 1.52s/62k/maybe	& 1.52s/62k/maybe	& 28.67s/151k/true	& 34.88s/151k/true \\ \cline{2-6}
	& bus+4 senders	& 10.94s/239k/maybe 	& 11.58s/239k/maybe    & 235.48s/765k/true	& 283s/766k/true\\ \hline
CSMA/CD	& bus+2 senders	& 0.05s/25k/true	& 0.06s/25k/true	& 0.06s/25k/true	& 0.72s/36k/true \\ \cline{2-6}
(C)	& bus+3 senders	& 0.14s/49k/true	& 0.21s/49k/true	& 0.29s/79k/true	& 5.51s/183k/true\\ \cline{2-6}
	& bus+4 senders	& 0.43s/97k/true	& 0.67s/97k/true	& 1.36s/298k/true	& 30.99s/752k/true\\ \cline{2-6}
	& bus+5 senders	& 1.32s/286k/true	& 2.44s/285k/true	& 6.73s/1045k/true	& 173.82s/2724k/true \\ \cline{2-6}
	& bus+6 senders	& 4.57s/833k/true	& 8.68s/835k/true	& 33.53s/3436k/true	& 907.41s/9031k/true \\ \cline{2-6}
	& bus+7 senders	& 16.32s/2364k/true	& 32.84s/2367k/true	& 166.14s/10652k/true	& 4558s/27993k/true \\ \hline
\end{tabular}
\end{center}
\caption{Performance w.r.t. non-Zeno requirements and EDGF techniques}
\label{tab.Z}
\end{table}
In general, we find that with or without EDGF technique, 
a non-Zeno requirement does add more complexity to the evaluation of the 
inevitability properties.  
Especially for the three specifications of CSMA/CD model, 
exponential blow-ups have been observed.  

For PATHOS benchmark, it is a totally different story.
The non-Zeno requirement seems to incur much less complexity than 
without it.  
After we have carefully traced the execution of our mode-checker, 
we found that this benchmark incurs very few iterations of loop (5) in {\tt gfp\_EDGF()}
although each iteration can be costly to run.  
On the other hand, it incurs a significant number of iterations of loop (6) in 
{\tt gfp\_Zeno\_EDGF()} although each iteration is not so costly.  
The accumulative effect of the loop iterations result in performance that 
contradicts our expectation.  
This benchmark shows that the evaluation performance of inevitability properties is very 
involved and depends on many factors.  

Futhermore, benchmark CSMA/CD (B) shows that some inevitability properties 
can only be verified with non-Zeno computations.  

As for the performance of the EDGF technique, we find that when the technique
fails, it only incurs a small overhead.  
When it succeeds, it significantly improves performance two to three-fold.

\subsection{Performance w.r.t. abstraction techniques \label{subsec.exp.tctla}}

In table~\ref{tab.A}, we report the performance data of our {\tt red} 4.1
with respect to our three abstraction techniques.
\begin{table}
\begin{center}
\scriptsize 
\begin{tabular}{|l|l||r|r|r|r|}
\hline
benchmarks  &concurrency& \multicolumn{1}{c|}{no abstraction}
			& \multicolumn{1}{c|}{Game-abs.}
			& \multicolumn{1}{c|}{Game-discrete-abs.}
			& \multicolumn{1}{c|}{Game-mag.-abs.}	\\\cline{3-6}
	&		& time/space/answer	& time/space/answer	& time/space/answer	& time/space/answer \\ \hline
pathos	& 2 procs.s	& 0.03s/7k/true		& 0.01s/7k/true		& 0.03s/7k/true		& 0.03s/7k/true	 \\ \cline{2-6}
	& 3 procs.s	& 0.08s/17k/true	& 0.11s/17k/maybe	& 0.09s/17k/true	& 0.1s/22k/true	 \\ \cline{2-6}
	& 4 procs.s	& 0.31s/42k/true   	& 0.36s/36k/maybe	& 0.37s/36k/true	& 0.78s/100k/true \\ \cline{2-6}
	& 5 procs.s	& 1.17s/114k/true  	& 1.16s/71k/maybe	& 1.2s/71k/true		& 8.55s/674k/true \\ \cline{2-6}
	& 6 procs.s	& 5.22s/314k/true	& 2.83s/114k/maybe	& 3.39s/114k/true	& 191.1s/6074k/true\\ \cline{2-6}
	& 7 procs.s	& 30.71s/942k/true 	& 6.66s/175k/maybe	& 8.62s/175k/true	& 6890s/62321k/true \\ \hline
leader	& 2 procs.s	& 0.04s/16k/true	& 0.03s/16k/true	& 0.03s/16k/true	& 0.02s/16k/true \\ \cline{2-6}
election& 3 procs.s	& 0.25s/84k/true	& 0.25s/84k/true	& 0.23s/84k/true	& 0.25s/84k/true \\ \cline{2-6}
	& 4 procs.s	& 1.54s/338k/true	& 1.53s/338k/true	& 1.54s/338k/true	& 1.52s/338k/true \\ \cline{2-6}
	& 5 procs.s	& 11.23s/1164k/true 	& 11.71s/1164k/true	& 11.38s/1164k/true 	& 11.39s/1164k/true \\ \cline{2-6}
	& 6 procs.s	& 110.9s/7992k/true	& 111.2s/7993k/true	& 110.8s/7993k/true	& 110.2s/7993k/true \\ \cline{2-6}
	& 7 procs.s	& 860.6s/42062k/true	& 854.8s/42123k/true	& 861.5s/42123k/true	& 867.7s/42123k/true \\ \hline
CSMA/CD	& bus+2 senders	& 0.33s/42k/true	& 0.33s/42k/true	& 0.29s/42k/true	& 0.33s/42k/true \\ \cline{2-6}
(A)	& bus+3 senders	& 3.09s/191k/true	& 3.35s/191k/maybe	& 1.33s/191k/true	& 3.35s/191k/maybe  \\ \cline{2-6}
	& bus+4 senders	& 26.23s/936k/true	& 9.57s/731k/maybe	& 4.79s/731k/true	& 9.57s/731k/maybe \\ \cline{2-6}
	& bus+5 senders	& 195.14s/4501k/true	& 29.89s/2529k/maybe	& 16.96s/2529k/true	& 29.8s/2529k/maybe\\ \hline
CSMA/CD	& bus+2 senders	& 1.92s/37k/true        & 0.58s/25k/true	& 0.76s/25k/true	& 0.58s/25k/true \\ \cline{2-6}
(B)	& bus+3 senders	& 28.67s/151k/true	& 2.73s/88k/true	& 3.91s/85k/true	& 2.73s/88k/true \\ \cline{2-6}
	& bus+4 senders	& 235.48s/765k/true     & 9.54s/290k/true	& 14.72s/281k/true	& 9.54s/290/true \\ \hline
CSMA/CD	& bus+2 senders	& 0.06s/25k/true        & 0.06s/25k/true     	& 0.05s/25k/true	& 0.06s/25k/true \\ \cline{2-6}
(C)	& bus+3 senders	& 0.29s/79k/true        & 0.19s/79k/true        & 0.18s/79k/true	& 0.19s/79k/true \\ \cline{2-6}
	& bus+4 senders	& 1.36s/298k/true       & 0.71s/298k/true	& 0.73s/298k/true	& 0.71s/298k/true \\ \cline{2-6}
	& bus+5 senders	& 6.73s/1045k/true	& 2.85s/1045k/true	& 2.90s/1045k/true	& 2.85s/1045k/true \\ \cline{2-6}
	& bus+6	senders & 33.53s/3436k/true	& 11.84s/3436k/true    	& 11.77s/3436k/true	& 11.84s/3436k/true \\ \cline{2-6}
	& bus+7 senders	& 166.14s/10652k/true	& 47.64s/10652k/true    & 47.84s/10652k/true    & 47.64s/10652k/true \\ \hline
\end{tabular}
\\All benchmarks run with non-Zeno requirement and EDGF on.
\end{center}
\caption{Performance w.r.t. abstraction techniques}
\label{tab.A}
\end{table}
In general, the abstraction techniques give us much better performance.  
Notably, game-discrete and game-magnitude abstractions seem to have 
enough accuracy to discern true properties.
  
It is somewhat surprising that 
the game-magnitude abstraction incurs excessive complexity for PATHOS benchmark.  
After carefully examining the traces generated by {\tt red}, 
we found that because non-magnitude constraints were eliminated, some of the 
inconsistent convex state-spaces in the representation became 
consistent. 
These spurious convex state-spaces make many more paths in our CRD and greately burden our greatest fixpoint calculation. 
For instance, the outer-loop (5) of procedure {\tt gfp\_EDGF()} takes 
two iterations to reach the fixpoint with the game-magnitude abstraction.  
It only takes one iteration to do so without the abstraction.  
In our previous experience, this abstraction has worked efficiently with reachability analysis. 
It seems that the performance of abstraction techniques for greatest fixpoint evaluation 
can be subtle.

\subsection{Performance w.r.t. Kronos}

In table~\ref{tab.K}, we report the performance of Kronos 5.2 w.r.t. the 
five benchmarks.  
\begin{table}
\begin{center}
\scriptsize 
\begin{tabular}{|l|l||r|r|r|r|}
\hline
benchmarks  & concurrency 	& \multicolumn{1}{c|}{no abstraction}
			& \multicolumn{1}{c|}{extrapolation}
			& \multicolumn{1}{c|}{inclusion}
			& \multicolumn{1}{c|}{convex-hull}	\\\cline{3-6}
	&		& time/space/answer& time/space/answer& time/space/answer& time/space/answer \\ \hline
pathos	& 2 procs	& 0.0s/true	& 0.0s/true	& 0.0s/true	& 0.0s/true	\\ \cline{2-6}
	& 3 procs	& 0.01s/true	& 0.01s/true	& 0.02s/true& 0.02s/true	\\ \cline{2-6}
	& 4 procs	& Q/N/C		& Q/N/C		& Q/N/C		& Q/N/C			\\ \hline
leader	& 2 procs	& 0.0s/true	& 0.0s/true	& 0.0s/true	& 0.0s/true	  \\ \cline{2-6}
election& 3 procs	& 0.01s/true 	& 0.01s/true	& 0.01s/true	& 0.01s/true	 \\ \cline{2-6}
	& 4 procs	& 0.05s/true	& 0.06s/true	& 0.04s/true	& 0.04s/true	 \\ \cline{2-6}
	& 5 procs	& Q/N/C		& Q/N/C		& Q/N/C		& Q/N/C			\\ \hline
CSMA/CD	& bus+2 senders	& 0.0s/true	& 0.01s/true	& 0.0s/true 	& 0.01s/true\\ \cline{2-6}
(A)	& bus+3 senders	& 0.01s/true 	& 0.01s/true	& 0.01s/true	& 0.01s/true\\ \cline{2-6}
	& bus+4 senders	& 0.06s/true 	& 0.06s/true	& 0.06s/true	& 0.06s/true\\ \cline{2-6}
	& bus+5 senders	& 0.31s/true	& 0.31s/true 	& 0.32s/true 	& 0.32s/true\\ \hline
CSMA/CD	& bus+2 senders	& 8.67s/true	& 8.68s/true	& 8.65s/true	& 8.71s/true \\ \cline{2-6}
(B)	& bus+3 senders	& O/M		& O/M		& O/M		& O/M	\\ \hline 
CSMA/CD	& bus+2 senders	& 2.69s/true	& 2.70s/true	& 2.72s/true	& 2.69s/true	\\ \cline{2-6}
(C)	& bus+3 senders	& O/M		& O/M		& O/M		& O/M	\\ \hline
\end{tabular}
\\Q/N/C means that Kronos cannot construct the quotient automata.   
\end{center}
\caption{Performance of Kronos in comparison}
\label{tab.K}
\end{table}
For PATHOS and leader election, Kronos did not succeed in constructing the quotient 
automata.  
But our {\tt red} seems to have no problem in this regard with its on-the-fly 
exploration of the state-space.  
Of course, the lack of high-level data-variables in Kronos' modeling language 
may also acerbate the problem.  

As for benchmark CSMA/CD (A), Kronos performs very well.  
We believe this is because this benchmark uses a bounded inevitability specification.  
Such properties have already been studied in the literature of Kronos\cite{Yovine97}.   

On the other hand, benchmarks CSMA/CD (B) and (C) use unbounded inevitability specifications
with modal-subformula nesting depths 1 and 2 respectively.  
Kronos does not scale up to the complexity of concurrency for these two benchmarks.  
Our {\tt red} prevails in these two benchmarks.

\section{Conclusion \label{sec.conc}}

How to enhance the performance of TCTL model-checking is a research 
issue to which people have not paid much attention. 
The reason may be that the reachability analysis problem for TA is already difficult 
enough and has absorbed much of our energy.  
Nevertheless, the issue is still important both theoretically and practically.  
Hopefully, this work can give us some insight to the complexity of the issue 
and attract more research interest in this regard. 
Specifically, the charts reported in subsection~\ref{subsec.d} may imply 
that further research is needed to investigate how to predict good $d$-values 
in real-world verification tasks.  
Our implementation shows that the ideas in this paper can be of potential use.

\newpage

\thispagestyle{empty}
\pagenumbering{roman}
\setcounter{page}{1}
\setcounter{section}{0}
\appendix
\noindent{\tt\LARGE APPENDICES}

\section{Procedure clock\_elimiante() \label{app.clock.eliminate}}

\hrule 
\vspace*{2mm} 
$\mbox{\tt clock\_eliminate}(\eta,x)$ \{
\hst	for each $x_1-x\sim c$ and $x-x_2\sim'c'$, if $\eta\wedge x_1-x\sim c\wedge x-x_2\sim'c'$ is not empty, \{
\hstt		$\eta_1:=\eta\wedge x_1-x\sim c\wedge x-x_2\sim'c$;
\hstt		$\eta:=\eta\wedge\neg \eta_1$;
		$\eta:=\eta\vee (\eta_1\wedge x_1-x_2\mbox{\tt compose\_upperbound}(\sim,c,\sim',c'))$;
\hst	\}
\hst	return $\eta$;
\\\}
\vspace*{2mm}
\\
$\mbox{\tt compse\_upperbound}(\sim,c,\sim',c')$ \{
\hst	if $c=\infty\vee c'=\infty$, return "$<\infty$"
\hst	else if $c=-\infty$,
\hstt		if $c'\leq 0$, return "$<-\infty$"; else return "$<-C_{A:\phi}+c'$;"
\hst	else if $c'=-\infty$,
\hstt		if $c\leq 0$, return "$<-\infty$"; else return "$<-C_{A:\phi}+c$;"
\hst	$c_r:=c+c'$;
\hst 	if $c_r>C_{A:\phi}\vee (\sim_r=\mbox{"$<$"}\vee c_r=C_{A:\phi})$,
		return "$<\infty$;"
\hst	if either $\sim$ or $\sim'$ is "$<$", $\sim_r$ is assigned "$<$",
	else $\sim_r$ is assigned "$\leq$"
\hst	else if $c_r<-C_{A:\phi}\vee(\sim_r=\mbox{"$<$"}\vee c_r=-C_{A:\phi})$,
		return "$<-\infty$;"
\hst 	else return "$\sim_r c_r$";
\\\}
\vspace*{2mm}
\hrule 

Procedure {\tt compose\_upperbound}() computes the new upperbound as the result
of adding two, up to the absolute bound of $C_{A:\phi}$.
For example, with $C_{A:\phi}=5$, $\mbox{\tt compose\_upperbound}(<,2,\leq, 3)=\mbox{"$<\infty$"}$
and $\mbox{\tt compose\_upperbound}(\leq ,2,\leq,1)=\mbox{"$\leq 3$"}$.

\section{Complete model-checking procedure with non-Zeno requirement \label{app.mck}}

\vspace*{3mm}
\hrule
\vspace*{2mm}
\noindent $\mbox{\tt model-check}(A,\phi)$ \{
\hst if $\eval(A,\emptyset,\neg\phi)$ is $\false$, return $\true$; else return $\false$.  
\\\} 
\\[2mm]
\noindent $\eval(A,\chi,\bar{\phi})$ \\
/* $\chi$ is the set of clocks declared in the scope of $\bar{\phi}$ */ \{ 
\hst switch ($\bar{\phi}$) \{
\hst {\bf case} ($\false$):
	return $\false$;
\hst {\bf case}  ($p$):
	return $p\wedge\bigwedge_{x\in \chi}x\geq 0$;
\hst {\bf case}  ($x-y\sim c$):
	return $x-y\sim c\wedge_{x\in \chi}x\geq 0$;
\hst {\bf case}  ($\phi_1\vee \phi_2$):
	return $\eval(A,\chi,\phi_1)\vee \eval(A,\chi,\phi_2)$;
\hst {\bf case}  ($\phi_1\wedge \phi_2$):
\hstt	return $\eval(A,\chi,\phi_1)\wedge \eval(A,\chi,\phi_2)$;
\hst {\bf case}  ($\neg \phi_1$):
	return $\neg \eval(A,\chi,\phi_1)$;
\hst {\bf case}  ($x.\phi_1$):
	return $\mbox{\tt clock\_eliminate}(x=0\wedge \eval(A,\chi\cup\{x\},\phi_1\wedge x\geq 0), x)$;
\hst {\bf case}  ($\exists \phi_1\until\phi_2$):
\hstt	$\eta_1:= \eval(A,\chi,\phi_1)$;
 	$\eta_2:= \eval(A,\chi,\phi_2)$;
\hstt	return $\breach(\eta_1,\eta_2)$;
\hst {\bf case}  ($\exists\pfrr\phi_1$): 
	return $\mbox{\tt gfp}(\eval(A, \phi_1))$;
\hst\}
\\\}
\vspace*{2mm}
\hrule
\vspace*{2mm}

\section{Abstract model-checking with \tctla \label{app.tctla}}

In the application of abstraction techniques, it is important to make them 
{\em safe}\cite{WongToi95}. 
That is to say, when the safe abstraction analyzer says a property is true, 
the property is indeed true.  
(But when it says false, we do not know whether the property is true.)   
There are two types of abstractions: 
{\em over-approximation} and {\em under-approximation}.  
The former means that the abstract state-space is a superset of the concrete
state-space, while the latter means that the abstract state-space is a subset of the 
concrete state-space. 
To make an abstraction safe means that we should stick to over-approximation
while evaluating $\exists\pfrr \neg\phi$ (the negation of the inevitability).  
But this can be difficult to enforce in general since negations 
deeply nested in formulas  
can turn over-approximation into under-approximation and thus make
abstractions unsafe.  

Usually inevitability properties do not occur on their own either.  
Instead, they are usually nested in other modal-formulas. 
For example, normally we may specify that 
\begin{center}
\em 
When there is a request, then eventually there is a service.
\end{center} 
In TCTL, this can be written as 
\begin{center}
\hfill 
$\forall \pfrr \left(\mbox{\tt request} \rightarrow \forall \pevt \mbox{\tt service}\right)$
\hfill 
$(f1)$ 
\end{center} 
In general, it can be difficult to restrict the negation of over-approximation 
from happening.  
But fortunately, formula $(f1)$ does not have such a problem. 
Consider the negation of formula $(f1)$, which is the following. 
\begin{center}
\hfill 
$\exists \pevt \left(\mbox{\tt request} \wedge \exists \pfrr \neg\mbox{\tt service}\right)$
\hfill 
$(f2)$ 
\end{center} 
Since there is no negation sign before any modal-formula, 
there is no problem with negation of over-approximation.  
Thus over-approximation can be applied here by doing over-approximations of 
the state sets thus satisfying the following subformulas in sequence (from left to right).
\begin{center} 
$\neg\mbox{\tt service}, \; 
\exists\pfrr\neg\mbox{\tt service}, \;
\mbox{\tt request}, \;
\mbox{\tt request}\wedge\exists\pfrr\neg\mbox{\tt service}, \;
(f2), \;
I\wedge (f2)$
\end{center}
If the state-set for $I\wedge (f2)$ is empty, then formula $(f1)$ is satisfied; 
else we do not have a conclusion.  
Note that the evaluation of state sets for 
$\neg\mbox{\tt service}$ and $\mbox{\tt request}$ respectively does 
not need any abstraction.  

The reasoning in the last paragraph can be extended to a general subclass of TCTL, 
$\tctla$.  
A formula is in \tctla iff the negation signs only appear before its atoms, and 
only universal path quantifications are used.  
For example, we may want to verify the following specification: 
\begin{center} 
$\forall \pfrr\left(\mbox{\tt request} 
	\rightarrow \forall \pfrr \left(\mbox{\tt service}
		\rightarrow \forall\pevt \mbox{\tt request}\right)\right)$  
\end{center} 
The formula says that if there is a request followed by a service, then 
from that service on, there will be a request.  
The negation of the specification is 
$\exists \pevt\left(\mbox{\tt request} 
	\wedge \exists \pevt \left(\mbox{\tt service}
		\wedge \exists\pfrr \neg\mbox{\tt request}\right)\right)$, 
which is in the subclass of \tctle, 
the subclass of TCTL with negations right before atoms and 
with only existential path quantifications.  
Note that the negations of formulas in $\tctla$ fall correctly in $\tctle$.   
The following lemma shows that over-approximation techniques 
with \tctle formulas always yield over-approximation. 

{\lemma: Given a \tctle formula $\phi$, 
if we evaluate each modal-subformula 
in $\phi$ with over-approximation, 
then we still get an over-approximation of the state set satisfying $\phi$.  
}
\\\pf 
This can be done by an inductive analysis on the structure of $\phi$.  
If $\phi$ is a literal expression of the forms $p$ or $\neg p$, then 
the evaluation does not involve any approximation. 
If $\phi$ is like $\phi_1\vee\phi_2$ or $\phi_1\wedge\phi_2$, 
then the evaluation of $\phi$ still yields over-approximation with 
the inductive hypothesis that the evaluations of $\phi_1$ and $\phi_2$ are both 
over-approximations. 
If $\phi$ is like $\exists \phi_1\until \phi_2$, then since 
the modal-formula is to be evaluated with over-approximation, with 
the inductive hypothesis, we know that  
$\phi$ is evaluated with over-approximation. 
The case for $\exists\pfrr\phi_1$ is similar.  
Thus the lemma is proven. 
\qed 

Our over-approximation technique is directly applied in procedure $\breach()$.  
In other words, we can extend $\breach()$ with over-approximation techniques as following.  
\begin{center}
$\breach^O(\eta_1,\eta_2)\equiv \mbox{\tt lfp} 
	Y.\mbox{\tt abs}\left(\eta_2\vee \left(\eta_1\wedge\mbox{\tt time\_bck}
	(\eta_1\wedge\bigvee_{e\in T}\mbox{\tt xtion\_bck}(Y, e))\right)\right)$.
\end{center}
Here $\mbox{\tt abs}()$ means a generic over-approximation procedure.  
Thus procedure $\breach^O()$ can be used in place of $\breach()$ 
in procedures {\tt gfp}(), {\evalg}(), and {\tt gfp\_EDGF}().  

In our tool {\tt red} 4.1, we have implemented a series 
of game-based abstraction procedures 
suitable for BDD-like data-structures and concurrent systems\cite{WHY03}.  
We use the term {\em "game"} here because we envision the concurrent system
operation as a game.
Those processes, which we want to verify, are treated as {\em players} while
the other processes are treated as {\em opponents}.
In the game, the players try to win (maintain the specification property)
under the worst (i.e., minimal) assumption on their opponents.
A process is a {\em player} iff its local variables appear in the
inevitability properties.
The other processes are called {\em opponents}.  
According to the well-observed discipline of modular programming\cite{Pressman82},
the behavioral correctness of a functional module should be
based on minimal assumption on the environment.
These {\em game-based abstraction} procedures omit opponents' state-information 
to make abstractions.  
\begin{list1}
\item {\em Game-abstraction}:
    The game abstraction procedure will eliminate the
    state information of the opponents from its argument state-predicate.
\item {\em Game-discrete-abstraction}:
    This abstraction procedure will eliminate all clock constraints for
    the opponents in the argument state-predicate.
\item {\em Game-magnitude-abstraction}:
    A clock constraint like $x-x'\sim c$ is called a {\em magnitude constraint}
    iff either $x$ or $x'$ is zero itself (i.e. the constraint is
    either $x\sim c$ or $-x'\sim c$).
    This abstraction procedure will erase all non-magnitude constraints
    of the opponents in the argument state-predicate.
\end{list1}
Details can be found in \cite{WHY03}.

\section{Benchmarks \label{app.benchmarks}} 

We use the following benchmarks to test our ideas and implementations. 
The specifications for the benchmarks fall in \tctla.  
Thus we can also carry out experiments with our abstraction techniques.  
\begin{list1}
\item {\em PATHOS real-time operating system scheduling specification}\cite{Balarin96}:\\
	In the system, each process runs with a distinct priority in a period equal 
	to the number of processes.
 	The biggest timing constant used is equal to the number of processes.
	The unbounded inevitability property we want to evaluate is that 
	"if the process with lowest priority is in the pending state, then inevitably 
	it will enter the running state thereafter." 
	For a system with three processes, this property is as follows. 
	\begin{center} 
	$\forall \pfrr \left(\mbox{\tt pending}_3
		\rightarrow\forall \pevt \mbox{\tt running}_3\right)$
	\end{center} 
	The nesting depth of the modal-operators is one.  
\item {\em Leader election specification}\cite{Wang03}: \\
	Each process has a local pointer {\tt parent} and a local clock.
	All processes initially come with its {\tt parent} = NULL.
	Then a process with its {\tt parent} = NULL may broadcast its request to be
	adopted by a parent.
	Another process with its {\tt parent} = NULL may respond.
	The process with the smaller identifier will become the parent of the
	other process in the requester-responder pair.
	The biggest timing constant used is $2$.
	The unbounded inevitability we want to verify is that eventually, the
	algorithm will finish with a unique leader elected.
	That is 
	\begin{center} 
	$\forall\pevt \left(\mbox{\tt parent}_1=\mbox{\tt NULL}\wedge
		\forall i:i\neq 1, \left(
			\mbox{\tt parent}_i\neq \mbox{\tt NULL}
			\wedge\mbox{\tt parent}_i<i
		\right)
	\right)
	$ 
	\end{center} 
	There is no nested modal-operators.  
	To guarantee the inevitability, 
	we assume that a process with {\tt parent} = NULL will
	finish an iteration of the algorithm in 2 time units.
\item {\em CSMA/CD protocol} \cite{Wang01a,Wang01b,Yovine97}:\\
	Basically, this is the ethernet bus arbitration protocol with
	the idea of collision-and-retry.
	The timing constants used are 26, 52, and 808.
	We have used three TCTL specifications for these benchmarks. 
	\begin{list2} 
	\item The first one requires that, when two processes are simultaneously 
		in the transmission mode, then in 26 time units, the bus will 
		inevitably go back to the idle state.  
		The property is a bounded inevitability and can be written as follows: 
		\begin{center} 
		\hfill 
		$\forall \pfrr \left(
			\left(\mbox{\tt transm}_1\wedge\mbox{\tt transm}_2\right)
				\rightarrow x.\forall \pevt\left(
					x<26\wedge \mbox{\tt bus\_idle}
				\right)
		\right)
		$
		\hfill 
		(A)
		\end{center} 
		Note that the inevitability is timed to happen in 26 time units.  
		This experiment allows us to observe how our techniques perform 
		with bounded inevitability.  
	\item The second specification requires that 
		if sender 1 is 
		in its {\tt transmission} mode for no less than
		52 time units, then it will inevitably enter the {\tt wait} mode.  
		\begin{center}
		\hfill
		$\forall \pfrr ((\mbox{\tt transm}_1\wedge x_1\geq 52)\rightarrow
			\forall \pevt \mbox{\tt wait}_1)$
		\hfill
		(B)
		\end{center}
		Specially, this specification can only be verified by quantifying only on non-Zeno computations.
	\item The third specification requires that
		if the bus is in the {\tt idle} mode and later enters the {\tt collision} mode,
		then it will inevitably go back to the {\tt idle} mode.
		This unbounded inevitability property is as follows.
		\begin{center}
		\hfill
		$\forall \pfrr \left(\mbox{\tt bus\_idle}
			\rightarrow \forall \pfrr \left(\mbox{\tt bus\_collision}
				\rightarrow \forall \pevt \mbox{\tt bus\_idle}
			\right)
		\right)$
		\hfill
		(C)
		\end{center}
		This property is special in that the nesting depth of modal-operator
		is two and can give us some insight on how our abstraction techniques
		scale to the inductive structure of specifications.
	\end{list2}
\end{list1}

\section{Performance data w.r.t. $d$-values \label{app.d.table}}

The performance data with various $d$-value setting in the evaluation of 
greatest fixpoint with non-Zeno computation requirement is in table~\ref{tab.d}.  
\begin{table}[h]
\begin{center}
\scriptsize
\begin{tabular}{|l|l||r|r|r|r|}
\hline
benchmarks  &$d$-values & \multicolumn{1}{c|}{2 procs}
			& \multicolumn{1}{c|}{3 procs}
			& \multicolumn{1}{c|}{4 procs}
			& \multicolumn{1}{c|}{5 procs}	\\\cline{3-6}
	&		& time/space	& time/space	& time/space	& time/space \\ \hline

leader	&$>$2	&0.03s/16k	&0.24s/84k	&1.58s/338k	&11.625s/1164k    \\ \cline{2-6}
	&$>$1	&0.04s/16k	&0.42s/89k	&3.55s/482k	&28.15s/1612k     \\ \cline{2-6}
	&$>=$1	&0.05s/16k	&0.53s/88k	&3.41s/451k	&30.52s/1554k     \\ \hline

patho	&$>$5	&\multicolumn{3}{l|}{not available}	 	&1.19s/114k	\\ \cline{2-2}\cline{5-6}
	&$>$4	&\multicolumn{2}{r|}{}		&0.32s/42k	&1.17s/109k     \\ \cline{2-2}\cline{4-6}
	&$>$3	&		&0.09s/17k	&0.31s/41k	&1.41s/119k     \\ \cline{2-6}
	&$>$2	&0.03s/7k	&0.08s/17k	&0.37s/51k	&1.83s/221k	\\ \cline{2-6}
	&$>$1	&0.02s/7k	&0.11s/21k	&0.55s/76k	&3.16s/244k     \\ \cline{2-6}
	&$>=$1	&0.04s/7k	&0.16s/26k	&0.65s/76k	&3.2s/245k      \\ \hline

CSMA/CD	&$>$808	&1.14s/45k	&93s/1418k	&2394.12s/10936k   &\\ \cline{2-5}
(A)	&$>$646	&0.94s/44k	&65.63s/1232k	&1616.22s/8581k    &\\ \cline{2-5}
	&$>$404	&0.72s/43k	&34.36s/877k	&636.53s/5336k     &\\ \cline{2-5}
	&$>$161	&0.54s/42k	&13.24s/501k	&185.17s/2539k     &\\ \cline{2-5}
	&$>$80	&0.45s/42k	&6.63s/308k	&65.41s/1483k      &\\ \cline{2-5}
	&$>=$64	&0.46s/42k	&5.3s/259k	&51.77s/1230k      &\\ \cline{2-5}
	&$>=$32	&0.34s/42k	&3.22s/191k	&27.16s/936k       &\\ \cline{2-5}
	&$>=$16	&0.70s/82k	&6.58s/426k	&55.5s/1640k       &\\ \cline{2-5}
	&$>=$8	&1.43s/96k	&14.5s/510k	&123.68s/1958k     &\\ \cline{2-5}
	&$>=$4	&2.66s/118k	&28.37s/639k	&255.59s/2446k     &\\ \cline{2-5}
	&$>=$2	&5.54s/165k	&62s/900k	&516.76s/3430k     &\\ \cline{2-5}
	&$>=$1	&15.64s/270k	&186.86s/1480k	&1443.97s/5587k    &\\ \hline

CSMA/CD	&$>$808	&0.66s/35k	&34s/719k	&863.29s/5198k     &\\ \cline{2-5}
(B)	&$>$646	&1.38s/62k	&48.45s/658k	&1095.99s/4187k    &\\ \cline{2-5}
	&$>$323	&1.15s/40k	&31.14s/421k	&507.73s/2178k     &\\ \cline{2-5}
	&$>$161	&1.27s/37k	&26.61s/265k	&320.67s/1199k     &\\ \cline{2-5}
	&$>$80	&1.91s/37k	&29.08s/151k	&240.77s/765k      &\\ \cline{2-5}
	&$>=$64	&2.22s/37k	&28.01s/130k	&240.44s/631k      &\\ \cline{2-5}
	&$>=$32	&3.51s/56k	&36.31s/228k	&272.32s/801k      &\\ \cline{2-5}
	&$>=$16	&6.45s/59k	&69.62s/251k	&521.32s/875k      &\\ \cline{2-5}
	&$>=$8	&13.88s/67k	&143.44s/280k	&1068s/958k        &\\ \cline{2-5}
	&$>=$4	&30.22s/80k	&309.2s/326k	&2257s/1087k       &\\ \cline{2-5}
	&$>=$2	&73.32s/113k	&717.8s/423k	&5165s/1349k       &\\ \cline{2-5}
	&$>=$1	&230s/214k	&2119s/667k	&13630s/1951k      &\\ \hline

CSMA/CD	&$>$808	&0.19s/25k	&1.75s/80k	&12.7s/300k        &\\ \cline{2-5}
(C)	&$>$646	&0.17s/25k	&1.48s/80k	&10.09s/299k       &\\ \cline{2-5}
	&$>$404	&0.13s/25k	&1s/80k		&6.52s/299k        &\\ \cline{2-5}
	&$>$161	&0.08s/25k	&0.6s/80k	&3.68s/299k        &\\ \cline{2-5}
	&$>$80	&0.07s/25k	&0.4s/80k	&2.12s/299k        &\\ \cline{2-5}
	&$>=$64	&0.08s/25k	&0.42s/80k	&2.27s/299k        &\\ \cline{2-5}
	&$>=$32	&0.07s/25k	&0.32s/79k	&1.55s/298k        &\\ \cline{2-5}
	&$>=$16	&0.11s/36k	&0.66s/192k	&3.12s/863k        &\\ \cline{2-5}
	&$>=$8	&0.22s/47k	&1.51s/256k	&7.96s/1056k       &\\ \cline{2-5}
	&$>=$4	&0.44s/64k	&3.16s/355k	&18.61s/1438k      &\\ \cline{2-5}
	&$>=$2	&1.03s/98k	&9.07s/556k	&53.9s/2208k       &\\ \cline{2-5}
	&$>=$1	&3.3s/178k	&37.11s/1001k	&230.32s/3892k     &\\ \hline

\end{tabular}
\end{center}
\caption{Performance w.r.t. $d$-values}
\label{tab.d}
\end{table}

\section{Space complexity charts w.r.t. $d$-values \label{app.d.mem}}

\begin{figure}
\begin{center}
\begin{tabular}{cc} 
\epsfig{width=70mm, file=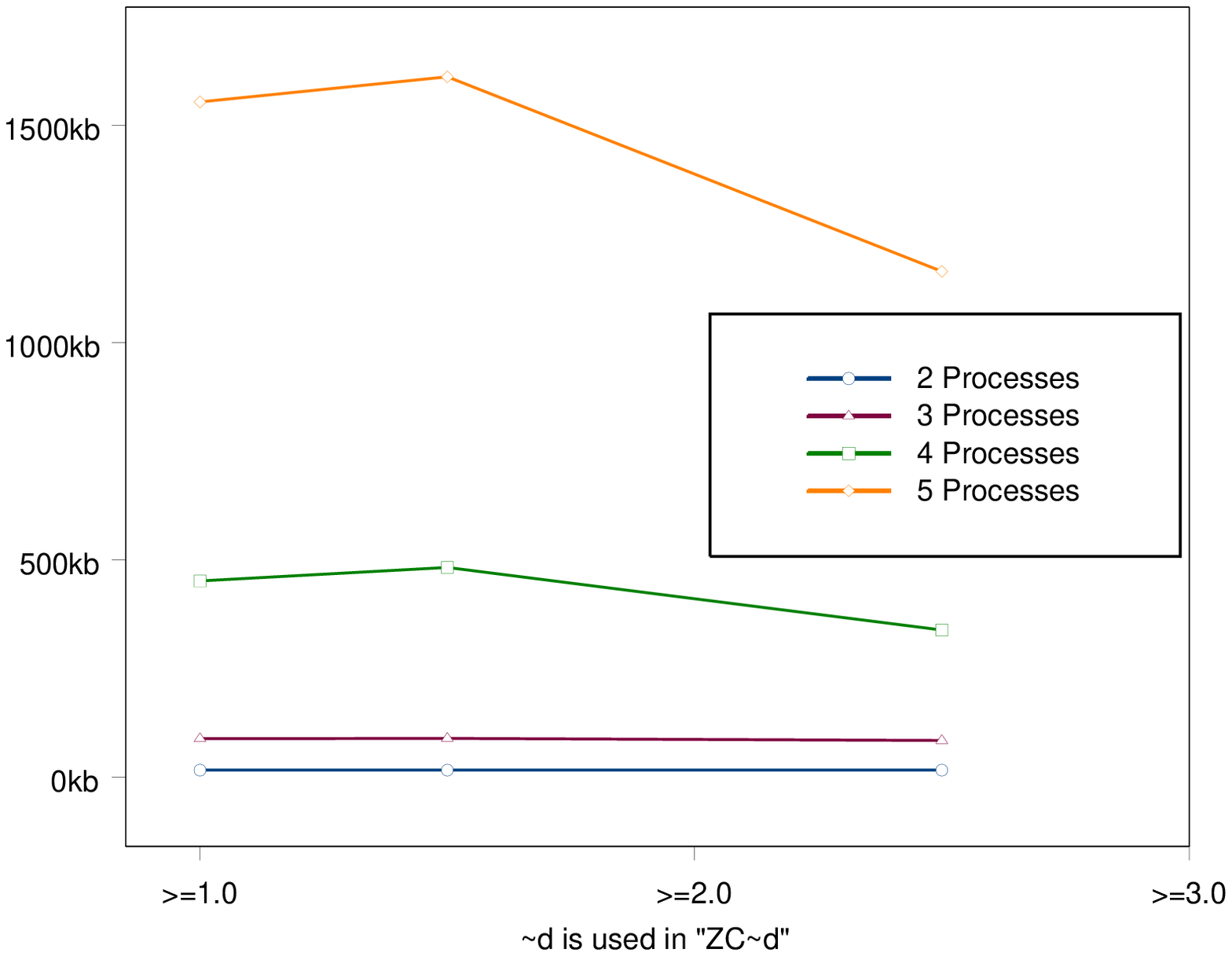} 
& \epsfig{width=70mm, file=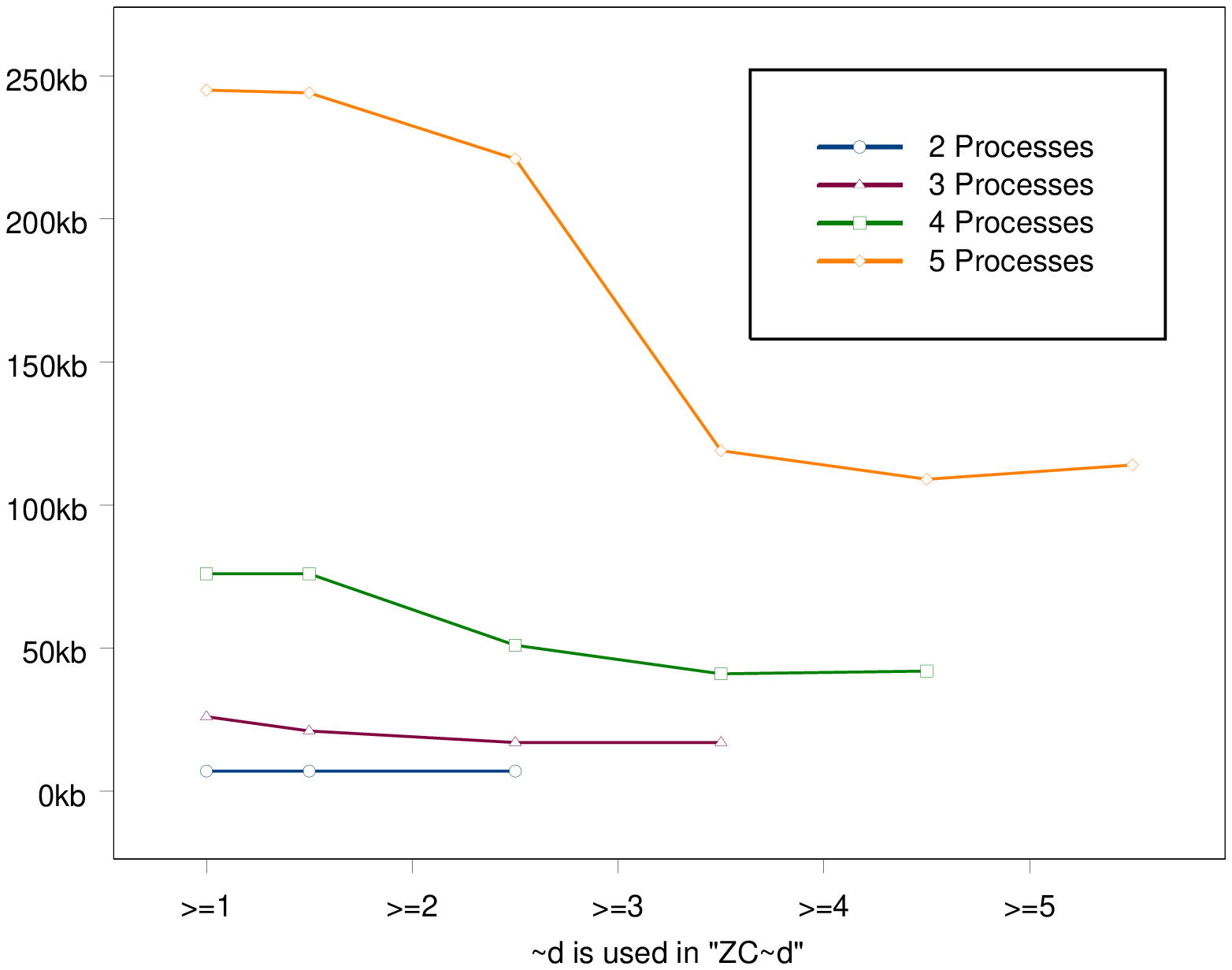} \\
(a) leader-election & (b) PATHOS \\
\epsfig{width=70mm, file=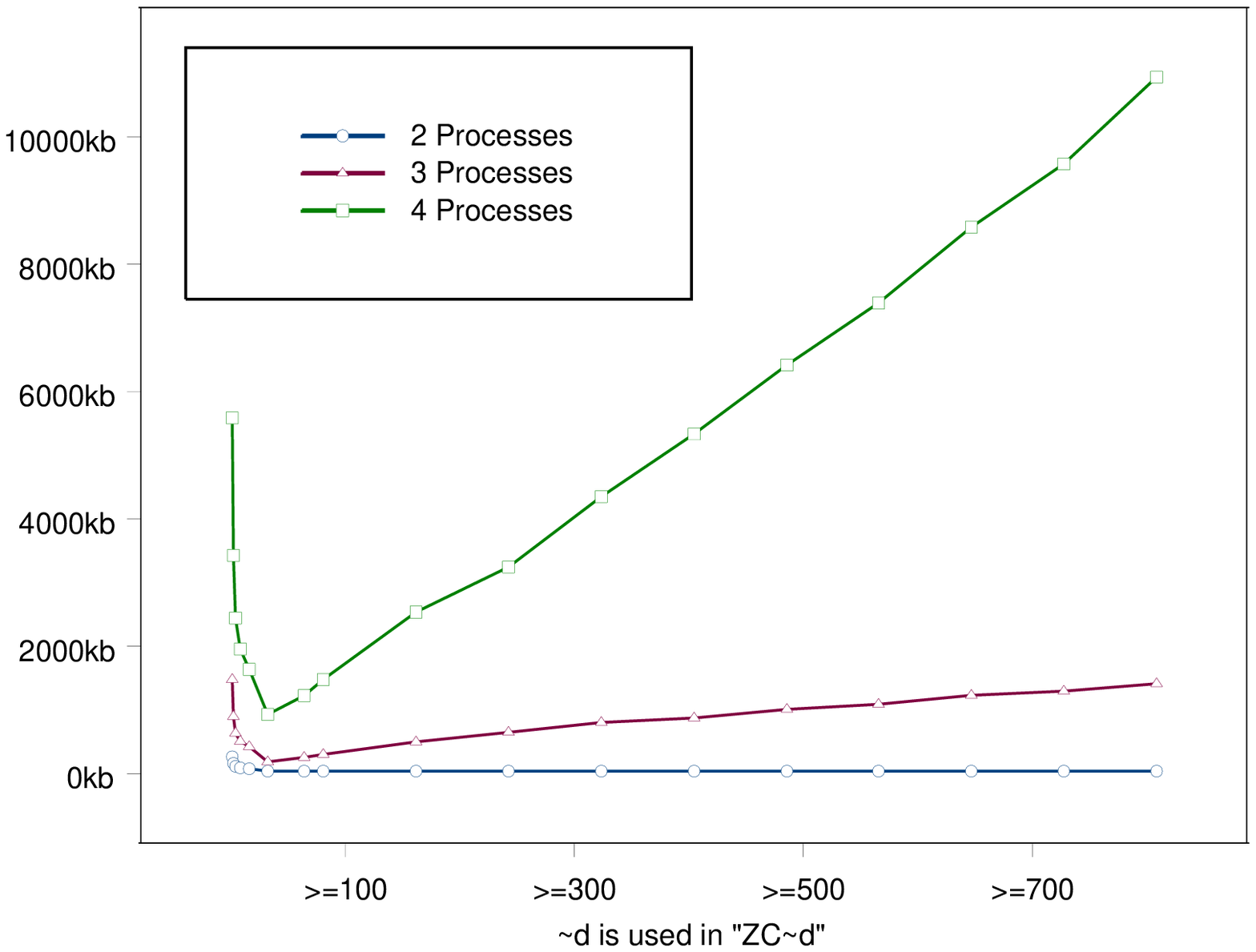}
& \epsfig{width=70mm, file=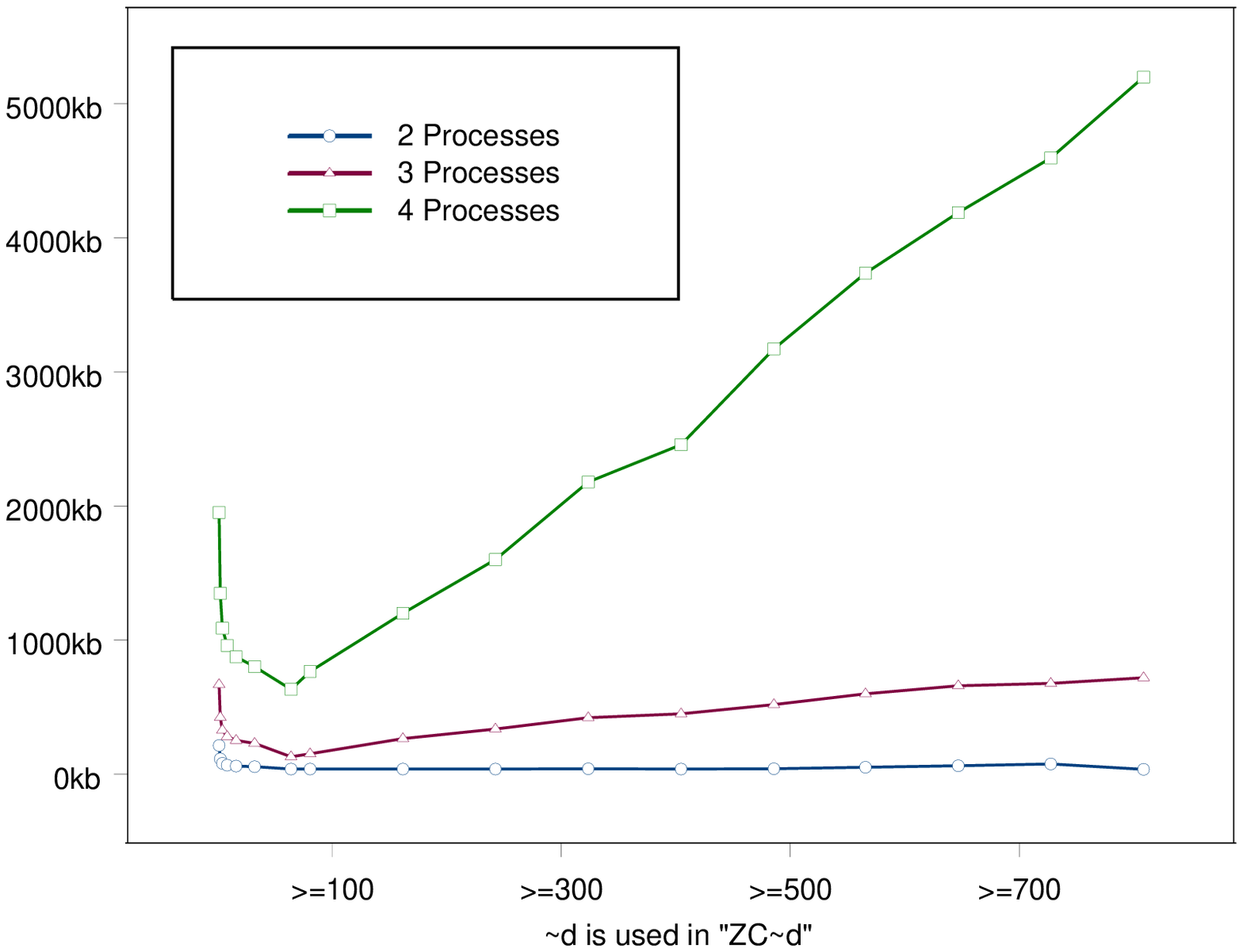}\\
(c) CSMA/CD(A) & (d) CSMA/CD(B) \\
\multicolumn{2}{c}{\epsfig{width=70mm, file=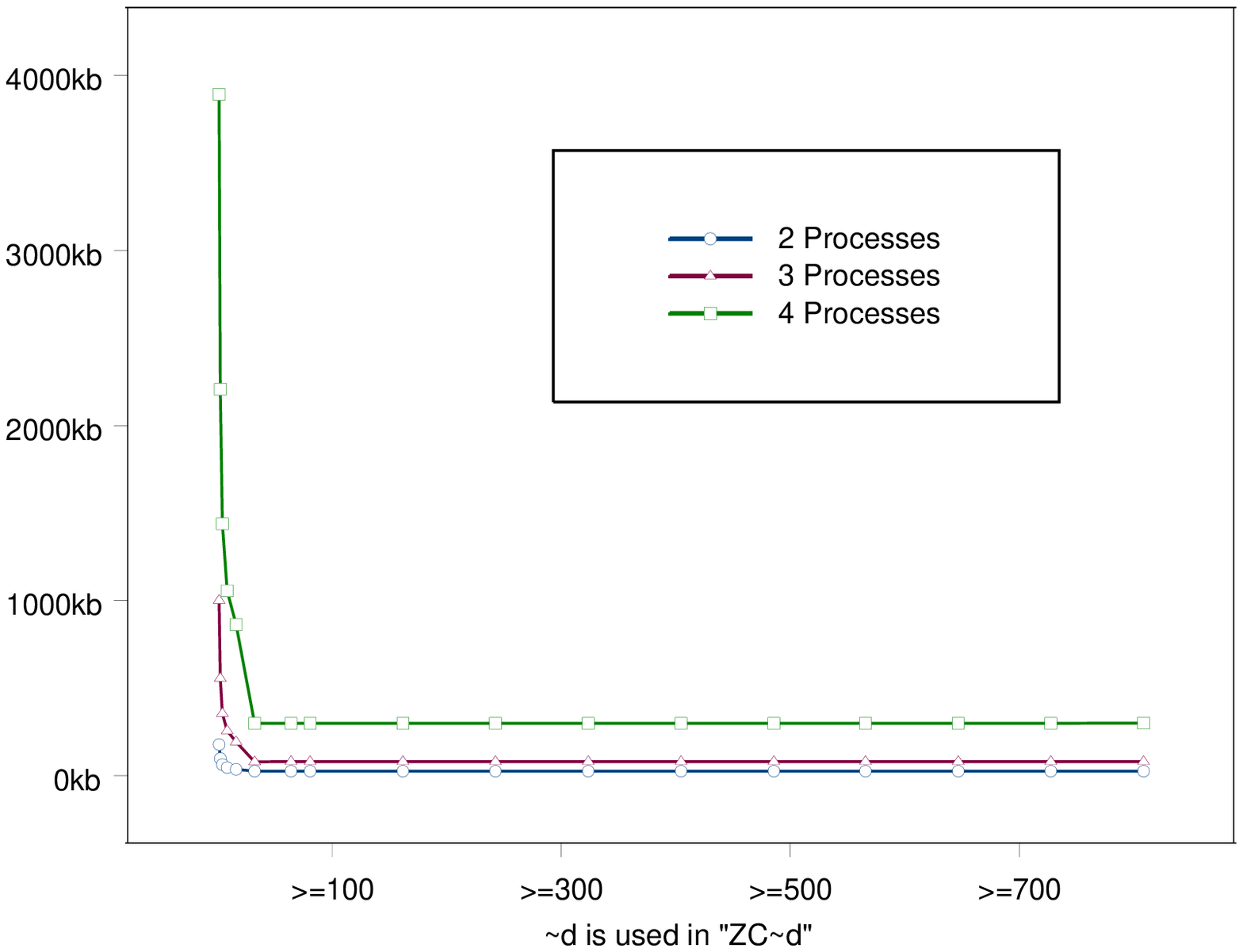}}\\
\multicolumn{2}{c}{(e) CSMA/CD(C)}
\end{tabular}
\end{center}
The Y-axis is with "memory space in kb"   
while the X-axis is with "$\sim d$" used in "$\mbox{\tt ZC}\sim d$."  
\caption{Memory-complexity charts w.r.t. $d$-values (Data collected with option EDGF)} 
\label{fig.charts.mem} 
\end{figure}

The charts for memory complexity with various $d$-parameter values is in 
figure~\ref{fig.charts.mem}.  

\end{document}